\documentclass[format=acmsmall, review=false]{acmart}

\usepackage{acm-ec-18}

\usepackage{booktabs} 

\usepackage[ruled]{algorithm2e} 

\SetAlFnt{\small}
\SetAlCapFnt{\small}
\SetAlCapNameFnt{\small}
\SetAlCapHSkip{0pt}
\IncMargin{-\parindent}

\usepackage{graphicx}  
\usepackage{tikz-qtree}
\usepackage{amsmath}
\usepackage{lipsum}
\usepackage{subcaption}
\usepackage{xcolor}

\newcommand{\leaveout}[1]{}

\usepackage[colorinlistoftodos]{todonotes}

\usepackage{algpseudocode}

\begin{document}

\title{Adversarial Coordination on Social Networks}
\author{Chen Hajaj}
\affiliation{
	\institution{Vanderbilt University}
	\city{Nashville}
	\state{TN}}
	
\author{Sixie Yu}

\affiliation{	\institution{Vanderbilt University}
	\city{Nashville}
	\state{TN}}
\author{Zlatko Joveski}
\affiliation{	\institution{Vanderbilt University}
	\city{Nashville}
	\state{TN}}
\author{Yevgeniy Vorobeychik}
\affiliation{	\institution{Vanderbilt University}
	\city{Nashville}
	\state{TN}}
\email{\{chen.hajaj,sixie.yu,zlatko.joveski,yevgeniy.vorobeychik\}@vanderbilt.edu}

\begin{abstract}
Decentralized coordination is one of the fundamental challenges for
societies and organizations.
While extensively explored from a variety of perspectives, one issue
which has received limited attention is human coordination in the presence
of adversarial agents.
We study this problem by situating human subjects as nodes on a
network, and endowing each with a role, either regular (with the goal
of achieving consensus among all regular players), or adversarial
(aiming to prevent consensus among regular players).
We show that adversarial nodes are, indeed, quite successful in
preventing consensus.
However, we demonstrate that having the ability to communicate among
network neighbors can considerably improve coordination success, as
well as resilience to adversarial nodes.
Our analysis of communication suggests that adversarial nodes attempt
to exploit this capability for their ends, but do so in a somewhat
limited way, perhaps to prevent regular nodes from recognizing their
intent.
In addition, we show that the presence of trusted nodes generally has
limited value, but does help when many adversarial nodes are present
and players can communicate.
Finally, we use experimental data to develop a computational model of
human behavior, and explore a number of additional parametric
variations, such as features of network topologies, using the
resulting data-driven agent-based model.
\end{abstract}
\maketitle


\section{Introduction}

Coordination is one of the fundamental problems faced by teams, organizations, and societies.
Such coordination problems are often decentralized, and involve limited local information and interaction, with such locality naturally captured by a network structure.
Considerable prior research has been devoted to understanding and modeling human behavior in networked coordination settings, such as networked consensus~\citep{Kearns2009,Judd2010,Kearns2012,vorobeychik2017does}, coloring~\citep{Mccubbins09,Judd2010}, bargaining~\citep{Chakraborty10}, and social dilemma games~\citep{Gracia12,Leibbrandt15}, among others.
However, decentralized coordination problems often take place in adversarial predicaments.
For example, organizations attempting to coordinate on a strategy may also compete with other organizations (legal and illicit), and coordination in combat mission planning and execution inherently faces adversarial entities in the form of enemy combatants.
Moreover, adversaries often attempt to exert their influence covertly, such as by bribing insiders, taking control of network nodes through cyber attacks, and spreading malicious influence tacitly through social networks, for example, by means of fake news.
Consequently, an important consideration in decentralized coordination is resilience to adversarial tampering with the process.
Surprisingly, this is not a question that has been studied to date from a behavioral perspective.

We investigate the problem of decentralized consensus on networks in the presence of adversarial nodes, leveraging two research modalities: human subject experiments and data-driven computational modeling.
Our experiments focus on two design factors: allowing neighboring nodes to communicate, and embedding a small set of trusted nodes in the network.
While communication has been a major subject of inquiry in prior research~\citep{Demichelis2008,Ellingsen2010,Miller2004,Cooper1992}, recent research suggests that communicating solely among network neighbors has limited impact on the ability of people to reach networked consensus~\citep{vorobeychik2017does}.
On the other hand, \citep{Abbas14} showed that the presence of trusted nodes can significantly facilitate decentralized coordination, albeit in a stylized model, rather with human subjects.
Our results run counter to both of these recent observations.
First, we demonstrate that communication helps a great deal, especially as we increase the number of adversarial nodes.
Second, we show that the presence of trusted nodes does not, in the aggregate, help; however, it \emph{is} of value when there is a substantial adversarial presence, and players can communicate.
Additionally, we observe that adversarial nodes clearly engage in deliberate attempts to manipulate outcomes.
For example, we observe that they tend to choose colors in opposition to their local neighborhood.
Moreover, when communication is enabled, they often send messages that are deliberately misleading. 
A surprising feature of adversarial behavior, however, is that their manipulation attempts are relatively subdued: their tendency to choose colors opposing local majority is relatively weak, and they rarely communicate in a way that blatantly disagrees with their objective local state.
We conjecture that this behavior is also partly strategic: since the identity of adversarial nodes is unobserved, remaining covert necessitates limiting the extent of malicious activity.

Our simulation experiments consider several additional analyses: 1) optimizing behavior models, through limited change to parameters, to maximize consensus rate, 2) optimizing location of trusted and adversarial nodes within the network, and 3) systematically considering the impact of parameters of network topology, such as density, clustering, and disparity in degree distribution, on consensus rates.
Overall, we observe that small changes in model parameters have little impact on consensus rates, and optimizing location is particularly beneficial for adversarial nodes, even when they do so following a similarly optimized placement of trusted players.
In addition, we find that network density improves consensus rate, with and without adversaries, but also increases the value of trusted nodes.
In contrast, clustering and disparity in degrees have limited impact, particularly when adversarial nodes are present.
\paragraph{Related Work}

Our study of networked coordination follows a number of prior efforts that investigate a variety of decentralized coordination problems on networks using human subjects methodology, such as \citep{Chakraborty10,Kearns2009,Judd2010,Kearns2012, Mccubbins09,vorobeychik2017does}.
The impact of communication on human coordination and cooperation has extensive, parallel literature, using both human subjects~\citep{Szamado2011,Richerson2010,Olmstead2009} and theoretical methods \citep{Farrell87,Farrell88,Demichelis2008,Ellingsen2010,Miller2004}.
However, in most of this literature, communication is grafted on as a distinct pre-play stage; moreover, much of this literature study simple, two-player games.
\citet{vorobeychik2017does} is a recent exception, combining both threads, but investigating only non-adversarial settings.

Resilience in coordination has been analyzed by several efforts, both theoretically and in simulations, using relatively stylized behavior models~\citep{Leblanc12,Leblanc13,zeng2014resilient}.
\citet{Abbas14} study the importance of trusted nodes in such settings, again, analyzing a stylized model of decentralized consensus, rather than human behavior.

Data-driven or empirical agent-based modeling have been proposed as a means of performing simulations that reliably reflect actual behavior data~\citep{Wunder13,Zhang16}.
Our simulation-based analysis follows in the spirit of these efforts.

Finally, our work has some relationship to experimental studies of Stackelberg security games~\citep{Pita10,Nguyen13}.
However, the specific context of these games is securing a collection of targets, using limited resources, rather than decentralized, networked coordination.

\section{Experimental Methodology}

\subsection{General Setup}

\begin{figure}[h!]
	\centering
	\includegraphics[width=4.5in]{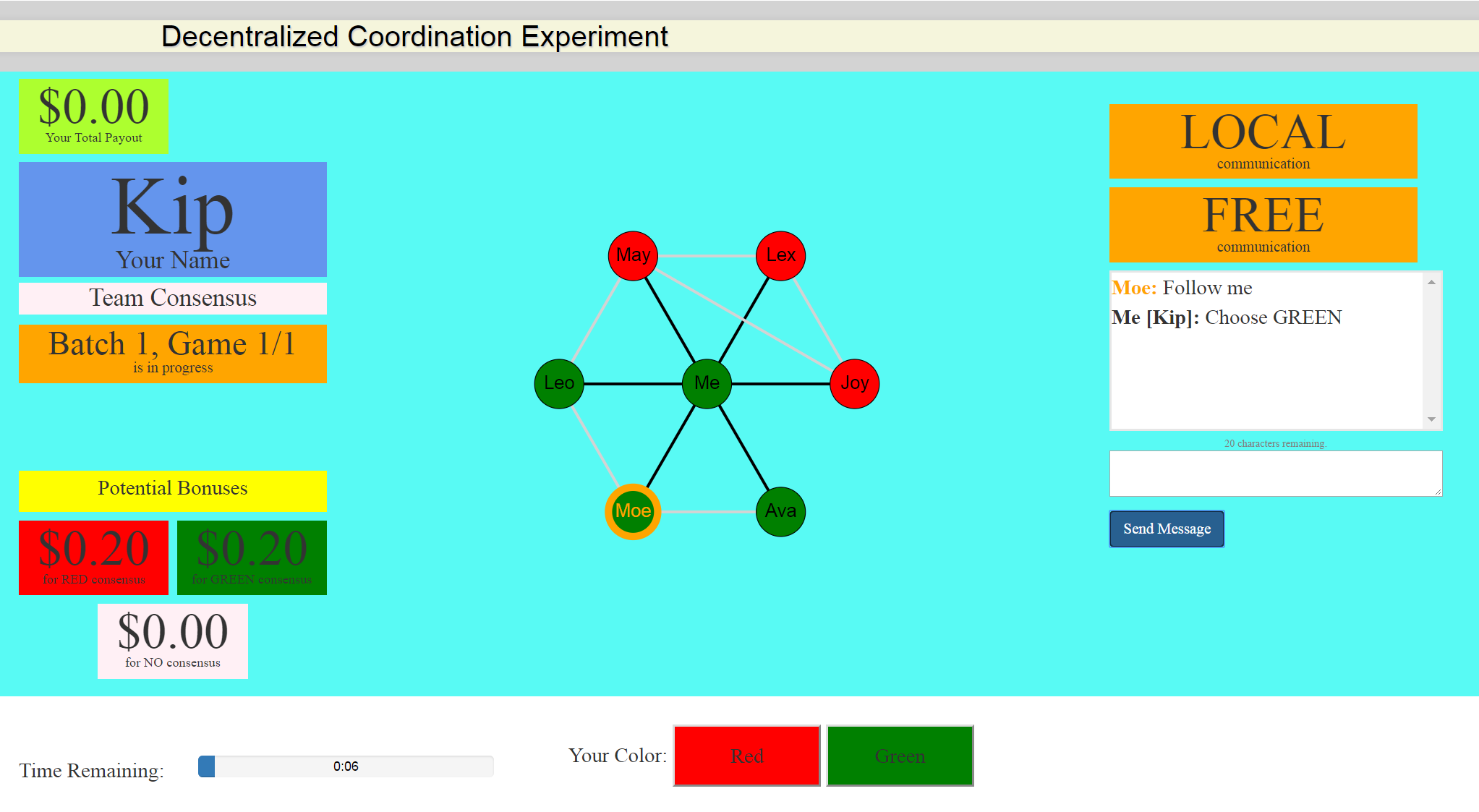}
	\begin{tabular}{ccc}
		\includegraphics[width=1.7in]{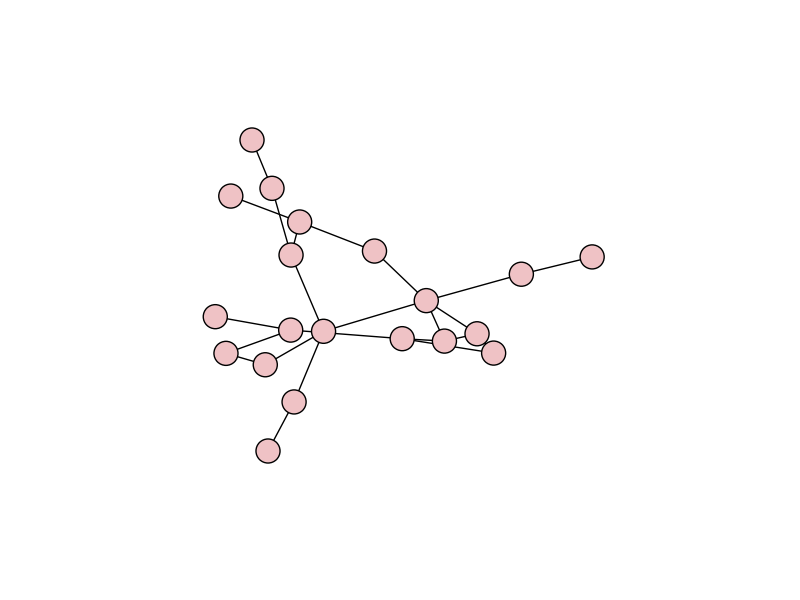} & 
		\includegraphics[width=1.7in]{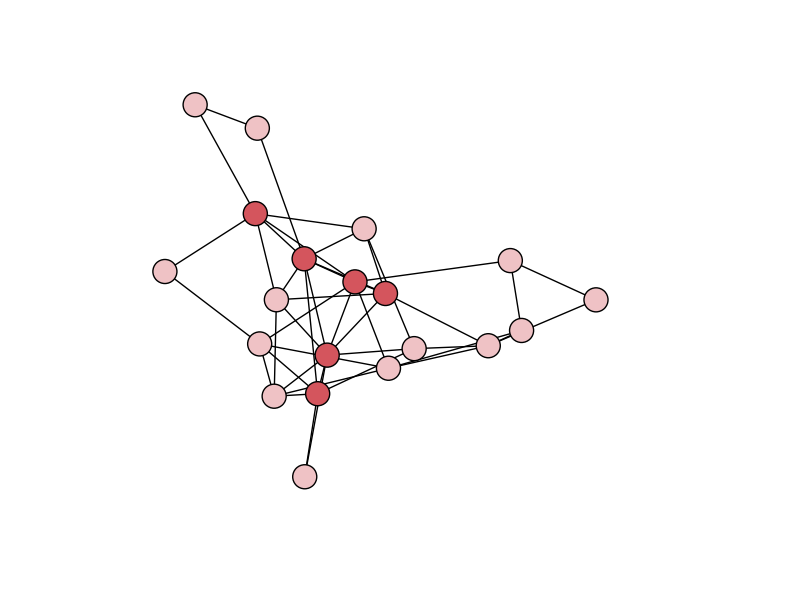} & 
		\includegraphics[width=1.7in]{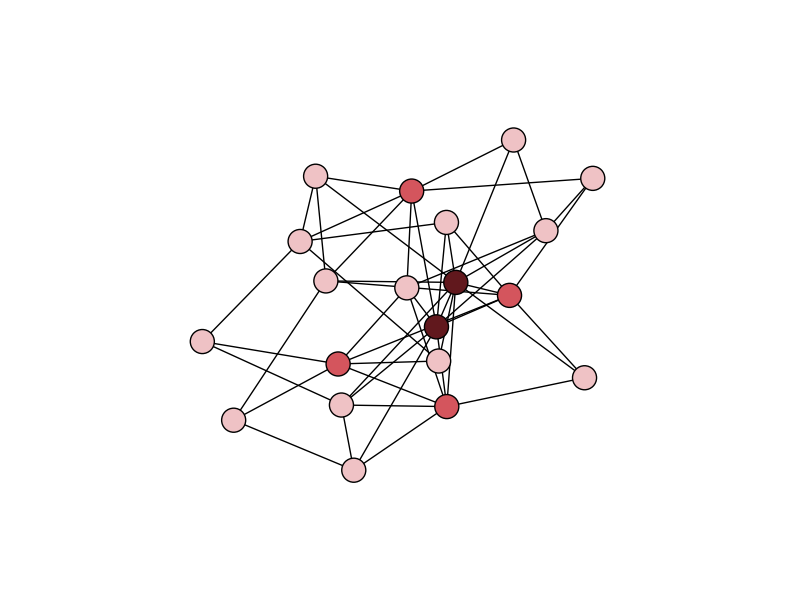}\\
		ER-Sparse & ER-Dense & BA
	\end{tabular}
	\caption{Top: an example graphical interface from the point of view of an experimental subject, who is represented by a node in the network.  The subject can see both her own node (labeled as ``Me'') and her network neighbors (labeled with their pseudonyms, randomly assigned at the beginning of a game), as well as connections among her neighbors.  The subject can also observe her current total payout in the experiment session (over all games played thus far).  
		Potential bonus, in case of consensus (either RED or GREEN), is shown on the left. Also on the bottom left portion of the interface, the subjects see the time remaining in the game.  Finally, in games involving communication, an instant-message-like interface is shown on the right, with a box where messages can be viewed and entered.  A clearly labeled sign describes whether the game involves (LOCAL) communication or no communication.  Bottom: example instances of networks, where darker colors indicate higher node degrees.}
	\label{F:gui}
\end{figure}

We designed a human subject experiment to study \emph{adversarial coordination on social networks}.
Specifically, the experiment builds on networked consensus games~\cite{Judd2010,kearns2012behavioral}, in which a collection of players (human subjects) act as nodes on an exogenously specified graph, choosing between two colors: RED and GREEN.
These games proceed for 60 seconds, with individuals able to make changes to their color choice in essentially real time.
Each player has an egocentric view of the game illustrated in Figure~\ref{F:gui}, where their node is displayed at the center, and their network neighbors are shown surrounding the ``Me'' node, along with their color choices, as well network connections among them.
Any node is displayed as white prior to actively choosing a color.
The display screen also shows time remaining in the game.
Each player receives a base payment for each game played (\$0.15), as well as a bonus of \$0.20 if a global consensus on either color is reached (the game ends as soon as consensus is reached).

The game description so far replicates features from all prior experiments in networked consensus.
A new feature, introduced by Vorobeychik et al.\cite{vorobeychik2017does}, allows network neighbors to communicate through an instant message-style interface, shown on the right in Figure~\ref{F:gui}.
To facilitate such communication, when allowed, each player is assigned a 3-letter name at the beginning of each game, and this name serves as their unique identifier in communicating with others.
Specifically, when a player sends a message through this interface, all their immediate network neighbors receive the message (this mode of communication was termed \emph{local} communication by Vorobeychik et al.\cite{vorobeychik2017does}).

We made one change to this general setup, which turns out to be quite consequential.
In all prior experiments, the interface featured a \emph{progress bar}, which shows how close the overall state is to global consensus (measured by the number of nodes disagreeing with majority color).
In our setting, however, such a progress bar communicates too much information, particularly when adversaries are present, and we consequently removed it (particularly since it doesn't have a clear motivation and was just a design artifact of prior experiments).
As we observe below, removing the progress bar increases the importance of communication, relative to findings reported by {vorobeychik2017does}.

\subsection{Design of Adversarial Consensus Games}

Starting with the basic experimental framework described above, we augment the experimental platform with several features in order to study how adversarial nodes impact the ability of the rest (i.e., the non-adversarial sub-network) to reach global consensus.
For this purpose, we divide players into two teams: a \emph{consensus} team and a \emph{no-consensus} team (in our parlance, these are \emph{adversaries}).
The goal of the \emph{consensus} team is to reach global consensus \emph{among members of this team only}, captured by the bonus payment structure described above.
The goal of the \emph{no-consensus} team is to prevent consensus among members of the \emph{consensus} team, which we incentivize by paying a \$0.40 bonus to members of this team if and only if consensus fails.
At the beginning of the game, each player is assigned to one of these teams, and this assignment is indicated in their view of the game (see left part of Figure~\ref{F:gui}).

We fixed the number of \emph{consensus} players in each game to 20, to control the baseline difficulty of the task (the underlying consensus problem on networks becomes more difficult as the network size grows, other things being equal).
In addition, we introduced in each game $a$ \emph{no-consensus} players, where $a \in \{0, 2, 5\}$.
The value of $a$ was not disclosed to the players at the beginning of a game; although an omniscient observer can infer it from the size of the network (which is 20 + $a$), no player could in fact, do this, since players could only observe their direct neighbors, and we limited the maximum degree to 15 to facilitate effective visualization.

A crucial part of our design was the invisibility of adversaries (no-consensus nodes) to others, including other adversaries, and vice versa.
On the other hand, it is often possible to have a small number of known \emph{reliable} or \emph{trusted} nodes on the network, for example, nodes which are particularly difficult to compromise due to a high amount of investment in their security, and such nodes can greatly facilitate consensus~\cite{Abbas14}.
To allow for this, we vary the number of \emph{visible} members of the \emph{consensus} team (henceforth, \emph{visible nodes}), $v \in \{0, 1, 2, 5\}$.
However, these nodes are visible only to their immediate network neighbors, highlighted by an orange circle around the corresponding nodes, as in Figure~\ref{F:gui} for the player with an assigned name ``Moe''.

\subsection{Network Topologies}

For each game, we exogenously specify a network topology, stochastically generated from one of three random graph models: two variations of Erdos-Renyi (ER) graphs~\cite{Erdos60}, and a Barabasi-Albert (BA, also known as preferential attachment) model~\cite{Barabasi99}.
The two variations of the ER model differ in network density: one we term ER-dense, and the other ER-sparse.
The 20-node version of the ER-dense model has average degree 5.1, while the ER-sparse networks have an average degree of 2.6.
BA networks have an average degree of 5.1 (same as ER-dense).
Average degrees increase slightly when we add adversarial nodes.
Figure~\ref{F:gui} shows example networks for each of the three network generative models.

\subsection{Recruiting and Scheduling}

We recruited subjects for the experiment using the Amazon Mechanical Turk (AMT) platform~\cite{paolacci2010running,mason2012conducting}, now in common use for economic experiments with human subjects~\cite{mason2012conducting,rao1989effect,peled2015study,hajaj2015improving,hajaj2017enhancing}.
Recruited subjects were directed to read detailed experiment instructions and consent to participate in the experiment (which was collected online).
Once we had a large enough pool of consented subjects, we scheduled experiment sessions.
For each experiment session, we recruited 30-35 subjects, to ensure that we have a sufficient number even when there are no-shows.
Upon arrival, subjects were placed in a waiting room, and if there were more subjects than nodes in a graph, they were randomly rotated each game.
Each session began with a series of 5 practice games, followed by 50-65 actual games in which we systematically varied 4 experimental variables:
\begin{enumerate}
	\item Number of adversaries (\emph{no-consensus} players): $a \in \{0, 2, 5\}$.
	\item number of visible nodes (within the \emph{consensus} team): $v \in \{0, 1, 2, 5\}$.
	\item network topology: ER-dense, ER-sparse, and BA.
	\item communication: allowed or not allowed.
\end{enumerate}
The full study protocol was approved by the Vanderbilt University IRB.
We recruited a total of 556 participants who jointly played 1080 games.


\section{Experimental Results}\label{sec:ExpResults}

We now analyze the results of the experiments.  
Throughout, we focus on consensus rate, or proportion of games reaching global consensus on a color, as a measure of coordination success.

\subsection{The Impact of Adversarial Players on Consensus Rate}

One would naturally expect that having adversarial players participate in the game would have a deleterious impact on consensus rate.
\begin{figure}[h!]
	\centering
	\begin{tabular}{cc}
		\includegraphics[width=0.4\linewidth]{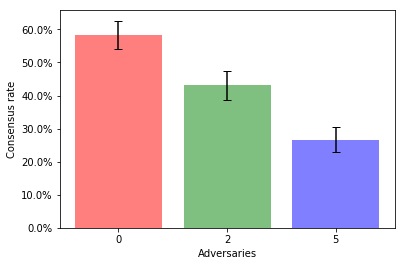} &
		\includegraphics[width=0.4\linewidth]{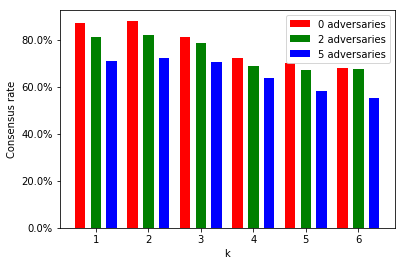} 
	\end{tabular}
	\caption{Impact of adversaries on the consensus rate.  Left: overall consensus rate, as function of the number of adversaries. Right: For each network distance, proportion of pairs of nodes with this distance between them who agree on a color at the end of the game.}
	\label{fig:AdvEffect}
\end{figure}
This intuition is readily confirmed in Figure~\ref{fig:AdvEffect} (left), with all differences statistically significant ($p<0.01$).
However, this observation obscures a crucial distinction between two kinds of impact adversaries can have in our setting:
\begin{enumerate}
	\item {\it Structural impact}: the adversarial nodes change network structure---in the extreme case, disconnecting the network among the \emph{consensus} team members, and
	\item {\it Behavioral impact}: behavior of adversarial nodes impacts the ability of the good nodes to reach consensus.
\end{enumerate}
There is a clear structural impact: 16\% of games with 2 adversaries, and 34\% of games with 5 adversaries become disconnected if we were to remove adversarial nodes.
In the cases in which adversarial nodes disconnect the graph, consensus rate drops to 14-15\%, roughly what one would expect by random chance (if we only have two connected components, and use the consensus rate of 58\% which is obtained with no adversaries for each component, the expected consensus rate is 17\%).
Of course, it is worth remembering that the network is not, in fact, disconnected, and adversarial nodes need to deliberately prevent the information about network state from spreading through them.
Indeed, not only do adversaries do so, the resulting consensus rates are slightly below expected, suggesting that adversarial behavior itself has an additional deleterious impact on the ability of nodes to coordinate.

To isolate the behavioral impact, in Figure~\ref{fig:AdvEffect} (right) we plot the proportion of times a pair which is $k$ network hops apart agrees on a color at the end of the game, as a function of network distance $k$ (we only include $k$ with at least 100 instances).
Here, we can still see a systematic decrease in coordination success, as a function of the number of adversaries, no matter how far apart nodes are.
For example, even network neighbors (i.e., $ k = 1 $) are finding it increasingly more difficult to agree on a color, on average, as we increase the number of adversaries.

\subsection{Communication Improves Resilience}

Next, we consider the impact that allowing players to communicate with their network neighbors has on their ability to coordinate successfully.
\begin{figure}[h!]
	\centering
	\begin{tabular}{cc}
		\includegraphics[width=0.4\linewidth]{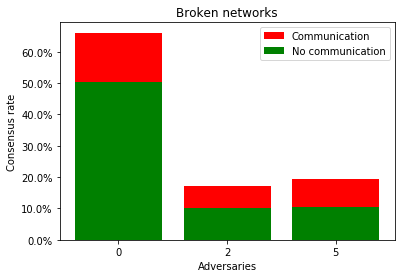}
		\includegraphics[width=0.4\linewidth]{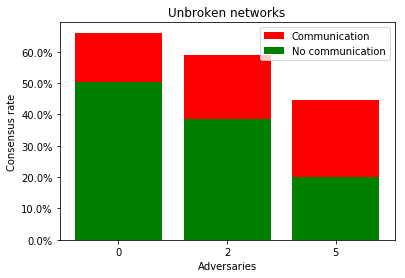}
	\end{tabular}
	\caption{The impact of communication on consensus rate. Left: Broken networks. Right: Unbroken networks.}
	\label{fig:comm}
\end{figure} 
Figure~\ref{fig:comm} shows that communication makes a clear impact (pooling broken and unbroken networks, all results are significant with $p < 0.01$).
In the aggregate, the value of communication increases with the number of adversaries: when no adversaries are present, communication increases consensus rate by $ 23.5\% $, with 2 adversaries improvement rises to $35.1\% $, and with 5 adversaries games that feature communication are $ 54.5\% $ more likely to reach consensus than those that don't.
The figure breaks these results into two plots: one when networks are broken if we were to remove adversarial node (left), and one for the remaining unbroken networks (right).
One would have expected that with broken networks consensus occurs largely by chance, and consequently, communication should have no impact.
We can observe that this is not so: even when networks are broken by adversaries, communication significantly increases consensus rate, nearly doubling it when there are 5 adversaries.

It is noteworthy that communication helps even when there are no adversaries, in contrast with the results reported by \citet{vorobeychik2017does}, who observe no significant difference in such cases and explain this by demonstrating that communication among network neighbors appears to be relatively uninformative.
The key distinction in our setting is the absence of progress bar: now that this source of global information is missing, communication becomes considerably more informative.

\begin{figure}[h!]
	\centering
	\begin{tabular}{ccc}
		\includegraphics[width=0.3\linewidth]{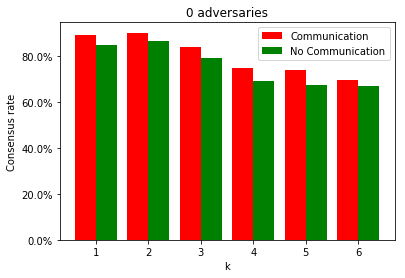} &
		\includegraphics[width=0.3\linewidth]{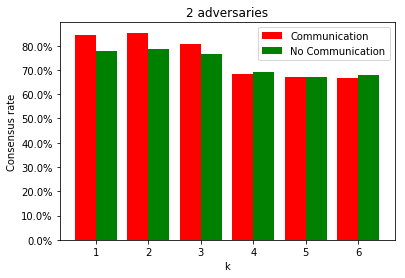} &
		\includegraphics[width=0.3\linewidth]{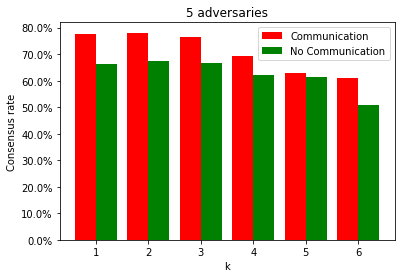}
	\end{tabular}
	\caption{The impact of communication on pairs of nodes agreeing in color choice, by node distance.}
	\label{fig:commbydist}
\end{figure}
Figure~\ref{fig:commbydist} unpacks the analysis of the impact of communication further by isolating, again, the behavioral impact of the adversaries, and the result is generally consistent, with communication increasing the likelihood a given pair of nodes agrees on a color at the end of the game, particularly when they are relatively close to each other in the network.

\subsection{The Impact of Network Structure}

Next, we consider what impact the network structure has on the ability of players to reach consensus with and without adversaries aiming to sabotage coordination.
\begin{figure}[h!]
	\centering
	\begin{subfigure}{0.4\linewidth}
		\includegraphics[width=1\linewidth]{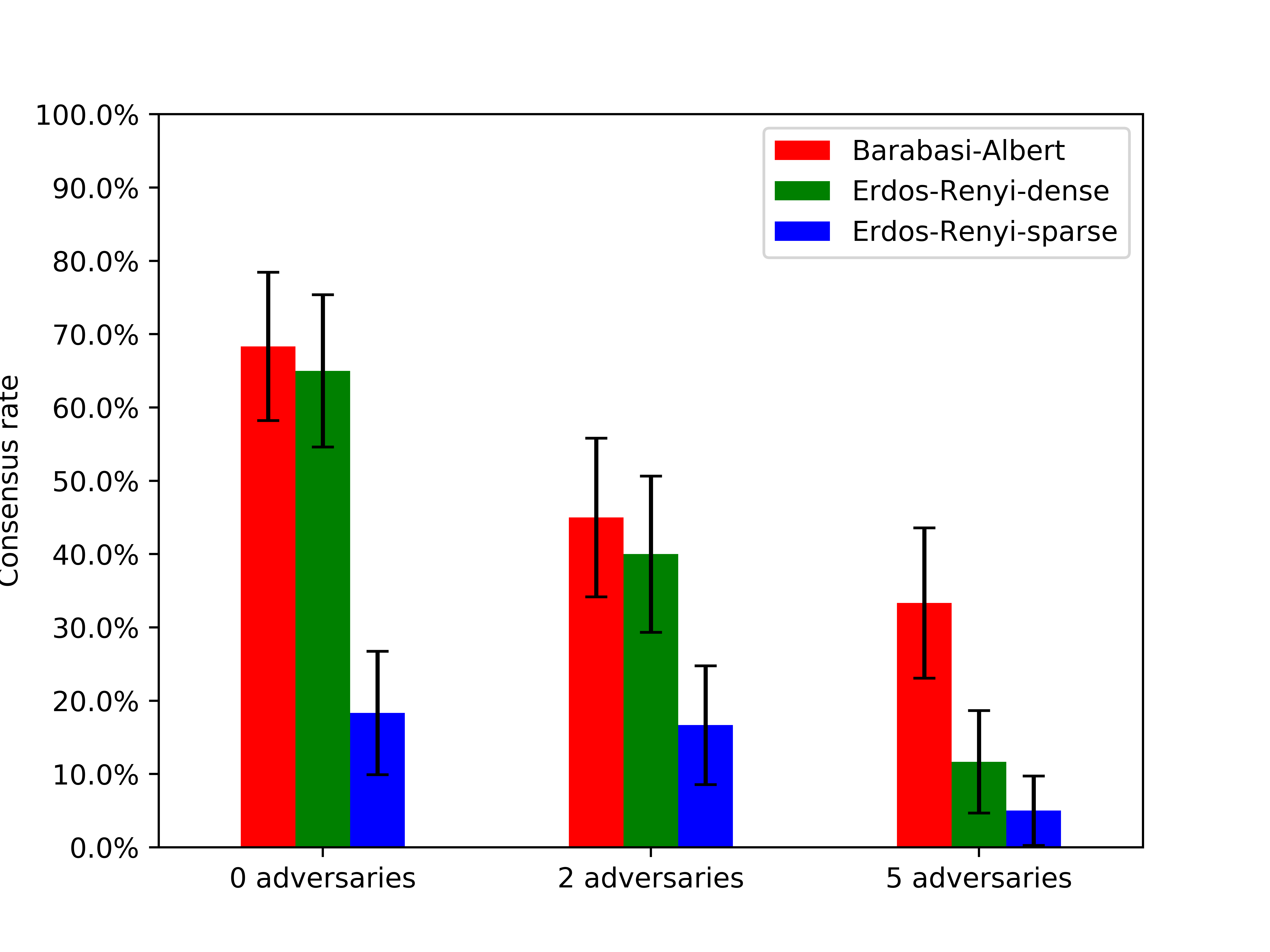}
		\label{fig:simAdv}
	\end{subfigure}%
	\begin{subfigure}{0.4\linewidth}
		\includegraphics[width=1\linewidth]{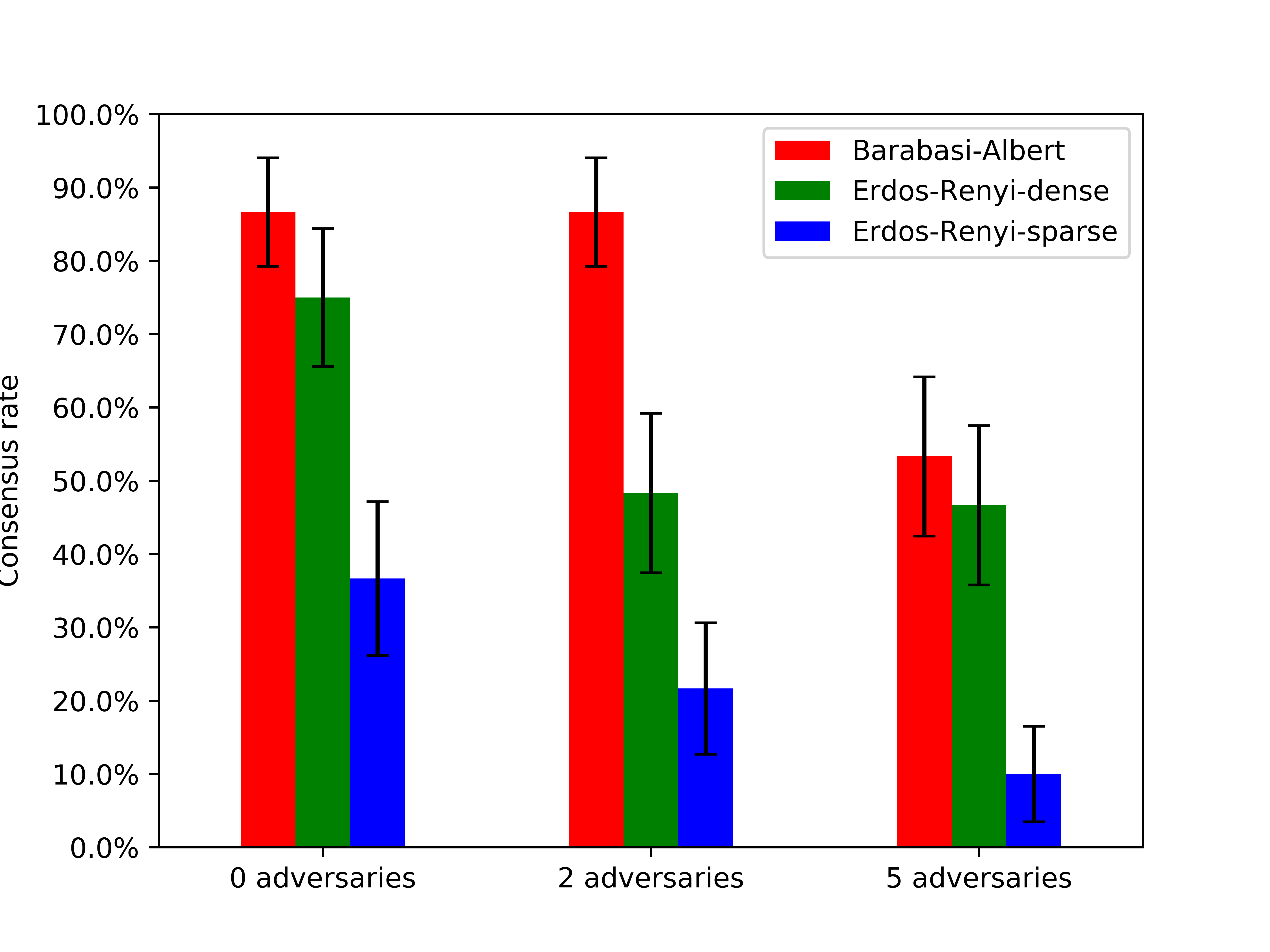}
		\label{fig:simNet}
	\end{subfigure}
	\caption{The effect of adversary players and network type on the consensus rate. Left: No communication. Right: Communication allowed.}
	\label{fig:AdvEffectType}
\end{figure}
Figure~\ref{fig:AdvEffectType} shows the results, broken up by network (BA, ER-dense, and ER-sparse), number of adversaries, and whether or not communication was allowed.
Perhaps the most dramatic impact that communication has is on BA networks: when communication is enabled, 2 adversaries are unable to significantly impact consensus rate, in contrast with games with no communication, where consensus rates of BA networks drop by over $30\%$.
This suggests that with few adversarial nodes, the ability to communicate endows BA networks with resilience \emph{even in the face of behavioral manipulation} by adversaries (which we observe to have a significant overall effect otherwise).
This finding complements the already well-known resilience of BA networks to random node removal~\citep{Albert00}.

\subsection{The Value of ``Trusted'' Nodes}

\begin{figure}[h!]
	\centering
	\includegraphics[width=0.4\linewidth]{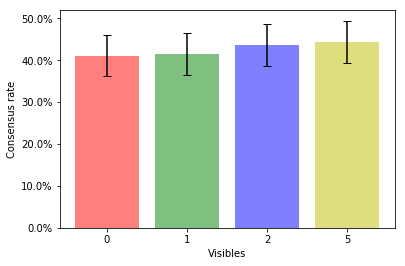}
	\caption{The effect of visible players on the consensus rate.}
	\label{fig:VisEffect}
\end{figure}
\citet{Abbas14} demonstrated that the presence of trusted nodes in a network can significantly improve resilience to attacks.
It is thus natural to hypothesize that nodes which are visible on the \emph{consensus} team (we can view these as trusted nodes, in the sense that they are known not to be adversarial) would significantly facilitate consensus.
Remarkably, Figure~\ref{fig:VisEffect} shows that this is not the case: as we increase the number of visible nodes, the impact on consensus rates is almost undetectable.
\begin{figure}[h!]
	\centering
	\begin{subfigure}{0.33\linewidth}
		\centering
		\includegraphics[width=1\linewidth]{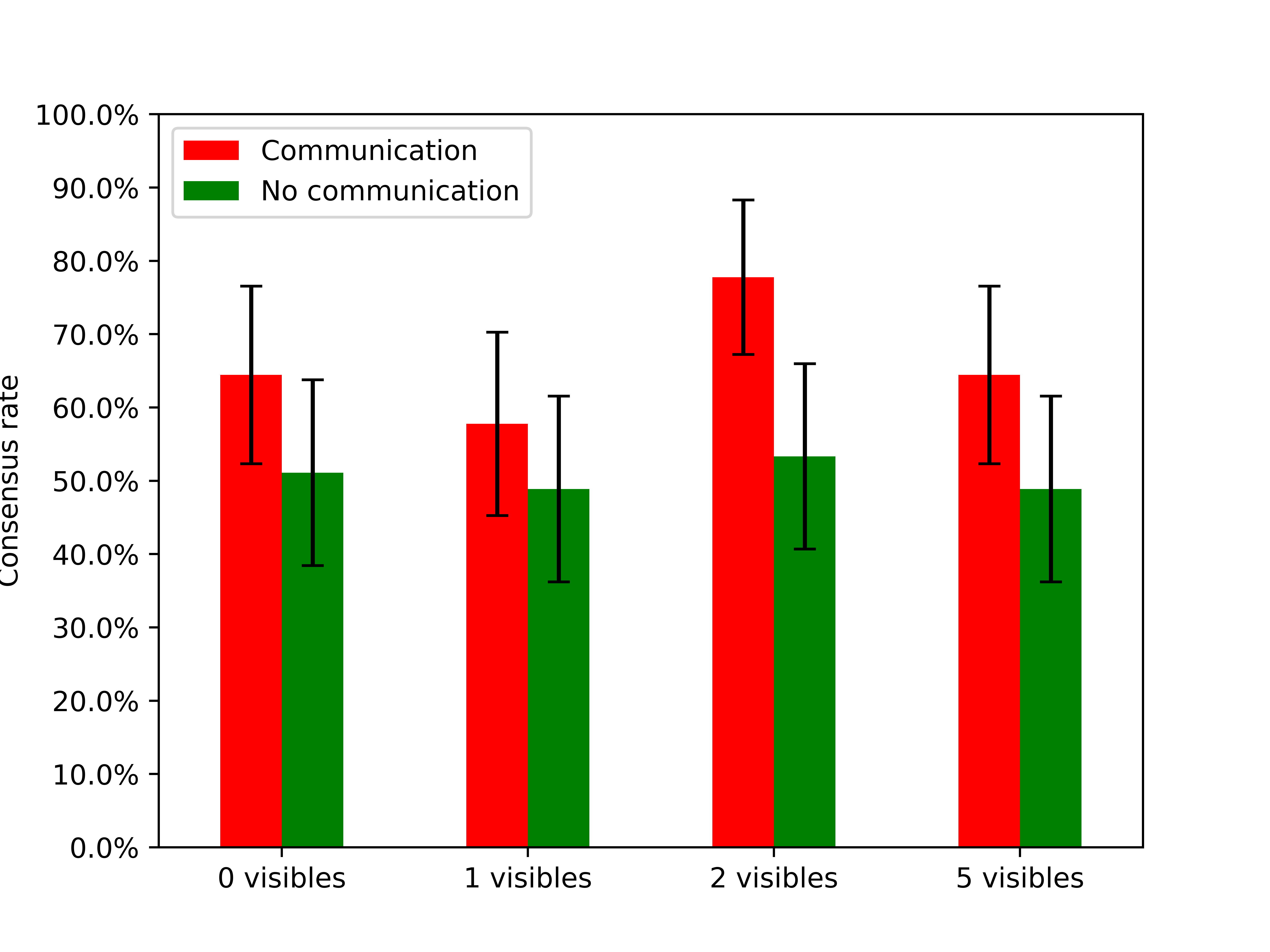}
		\caption{$ 0 $ adversaries}
		\label{subfig:0}
	\end{subfigure}%
	~
	\begin{subfigure}{0.33\linewidth}
		\centering
		\includegraphics[width=1\linewidth]{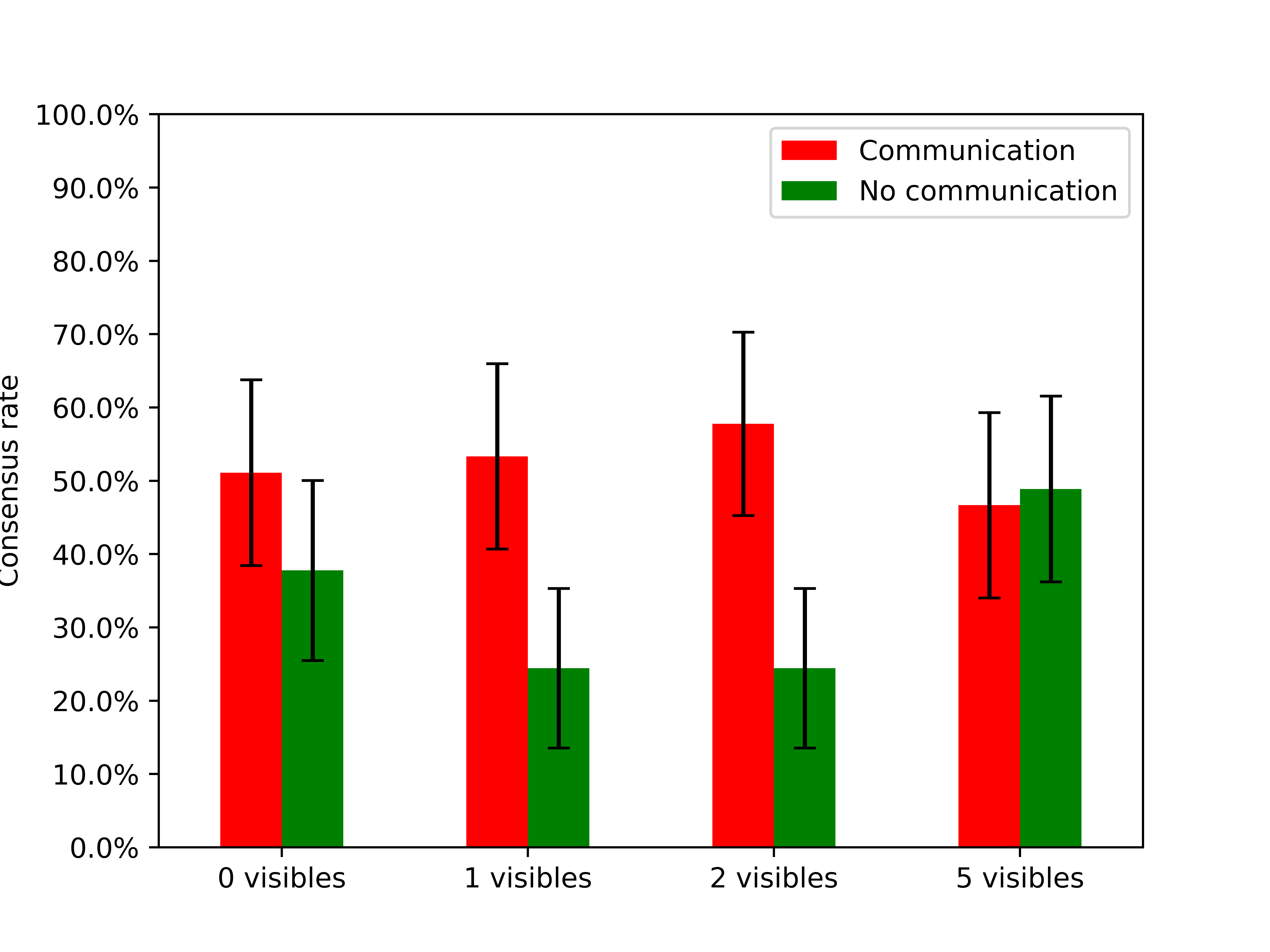}
		\caption{$ 2 $ adversaries}
		\label{subfig:2}
	\end{subfigure}%
	\begin{subfigure}{0.33\linewidth}
		\centering
		\includegraphics[width=1\linewidth]{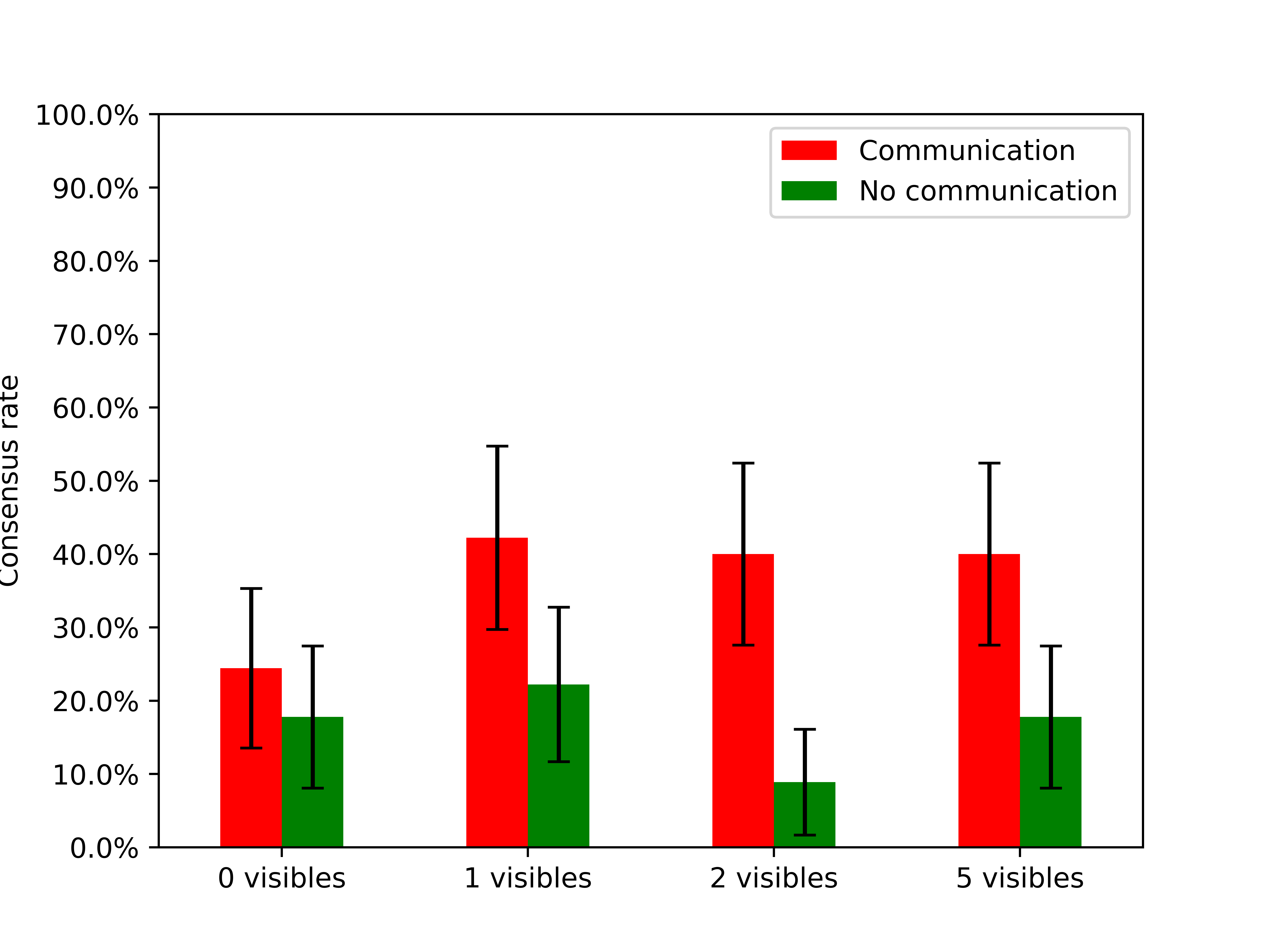}
		\caption{$ 5 $ adversaries}
		\label{subfig:5}
	\end{subfigure}%
	\caption{The effect of visible and adversarial players given the type of communication on the consensus rate.}
	\label{fig:CommEffectAdv}
\end{figure} 
To understand the impact of visible (trusted) nodes in greater depth, we unpack the results in Figure~\ref{fig:CommEffectAdv} by the number of visible nodes, the number of adversaries, and whether or not communication is allowed.
With 0 or 2 adversaries, there is a little systematic pattern to increase the number of visible nodes, whether or not communication is allowed.
There is, however, a signal when we have 5 adversaries and players can communicate: in this case, having at least 2 visible nodes is significantly better than having none ($p < 0.05$).
Thus, merely having trusted nodes is of dubious value, but allowing players (as well as the trusted node) to communicate can improve resilience when there are many adversarial nodes.

\section{Modeling and Analysis of Individual Behavior} 
\label{S:individual}

\subsection{Choosing and Changing Color}

We begin the analysis of individual behavior by observing an important difference in the frequency with which nodes with different roles change their color, on average.
We find that adversarial nodes change color significantly more often than others: 2.9 times per game, in comparison with visible consensus team players, who make only 2.1 changes in a game, and non-visible nodes, who change their color only twice a game, on average.
Despite these differences, it is also clear that players tend to change their color quite infrequently, generally well below once every 20 seconds.

To dig deeper into the nature of individual behavior, we now develop computational behavioral models, learned from the data collected in the experiments.
The purpose of these models will be two-fold: first, a better descriptive understanding of the behavior, and second, a data-driven agent-based modeling analysis that we discuss in Section~\ref{S:ddabm} below.

There are several complications in modeling human behavior in our settings.
The first is the fact that individuals may have three distinct roles:
\begin{enumerate}
	\item Adversarial node: a member of the \emph{no-consensus} team, whose goal is to prevent consensus among the ``good'' nodes (i.e., nodes on the \emph{consensus} team),
	\item Visible (``trusted'') node: a member of the \emph{consensus} team who is visibly a member of this team (that is, all neighbors can see that this node is on the \emph{consensus} team), and
	\item Regular nodes: all other members of the \emph{consensus} team.
\end{enumerate}
The second is that nodes, regular or not, may behave differently depending on whether they see visible nodes among their neighbors.
The third is the fundamental challenge of how we should model real-time color choices by the players.

To address the first two challenges, we created distinct behavioral models for the three roles, and distinct models for the situations when they have a visible node as a neighbor, and when they don't (thus, 6 models altogether).

To model behavior for any of these cases, we split the decision process into two qualitatively different parts: 1) the decision to choose the initial color, and 2) the decision to switch to a different color.
The rationale is that the initial decision is a deliberate choice of a particular color, and includes both the timing of changing from the initial default ``white'' color to either red or green, as well as the particular choice between these two.
In contrast, once a color is chosen, we expect a considerable amount of inertia to take hold, with players changing their color choices relatively infrequently.
Thus, modeling the decision to switch (or, effectively, the \emph{timing} of a color switch) naturally captures such inertia.
Finally, the initial decision was itself split into two models: the first modeling the timing of the initial color choice, and the second modeling which color is actually chosen.
Consequently, altogether we learned 18 different behavior models, or 3 models for each of the 6 roles and neighborhood assignments.
Next, we describe these 3 models (which are qualitatively the same for each of the role x neighborhood predicaments): \emph{timing of initial color choice}, \emph{choosing the initial color}, and \emph{timing of color change}.
All three use discrete time, discretized into 1-second intervals (so that the game lasts 60 discrete rounds).

\paragraph{\textbf{Timing of Initial Color Choice}}

Our first set of models predicts the timing of the initial choice of color, or, more precisely, the probability that the initial color is chosen in a discrete time unit prior to the first such choice.
For these models, the decision variables are:  $ D_{inv} $, the absolute difference between the fraction of a player's neighbors with unobserved team membership (\emph{consensus} or \emph{no-consensus})
that picked $red$ and the fraction that picked $green$; $D_{vis}$, the absolute difference between the fraction of a player's visible neighbors that picked $red$ and the fraction of those who picked  $green$ (if the player has visible neighbors); $N_{vis}$, the number of a player's neighbors that are visible, and $N_{inv}$, the number of a player's neighbors whose team membership is not observed.
(Note that $N_{vis} + N_{inv}$ is the total number of neighbors the player has).
The decision model is represented by a logistic regression with these coefficients, which we learned from experimental data.
We added $l_1$ (sparse) regularization to control for overfitting, with regularization parameter tuned using cross-validation.
All variables were normalized.

\begin{table}[h!]
	\caption{Color-picking model, P(pick a color).}
	\label{table:parameters_color_choosing_1}       
	\centering
	\begin{tabular}{cccccccc} 
		\hline\noalign{\smallskip}
		Type & Visible neighbors & Intercept & $D_{inv}$ & $D_{vis}$ & $N_{inv}$ & $N_{vis}$\\
		\noalign{\smallskip}\hline\noalign{\smallskip}
		Regular & No & $-1.952$ & $1.29$ & &  & \\
		& Yes & $-2.21$ & $0.548$ & $0.933$ & $ 0.002 $& $0.016$\\
		Visible & No & $-2.045$ & $1.742$ & & 0.04& \\
		& Yes & $-1.734$ & $0.579$ & $0.84$ & -0.061& 0.048\\
		Adversarial & No & $-2.284$ & $1.25$ & & 0.011&\\
		& Yes & $-2.744$ & $0.802$ & $0.662$ & 0.025& $0.155$\\
		\noalign{\smallskip}\hline
	\end{tabular}
\end{table}
The learned model coefficients for both the model with and without visible neighbors are given in Table~\ref{table:parameters_color_choosing_1}.
The results offer several interesting insights.
First, we can see that disagreement among neighbors stimulates a player to make an initial color choice earlier.
This is somewhat surprising, as we may expect players to wait until their neighbors had come to a near-consensus before making an initial move.
Second, disagreement among visible nodes has a more significant, positive impact on the likelihood of choosing a color at a particular time point.
Third, the behavior of adversarial nodes is broadly consistent with the first observation, but not with the second: such players appear to be more stimulated by disagreement among non-visible (i.e., those with unobserved team membership) than among visible neighbors.

\paragraph{\textbf{Choosing the Initial Color}}

Conditional on deciding to choose the initial color in a particular discrete time unit (per our previous models), the next decision we model is which of the two colors the player chooses.
We again use $l_1$-regularized logistic regression, where we predict the probability that a player chooses ``red'' as their initial color (conditional on choosing \emph{some initial color}).
As before, we use cross-validation to tune the regularization coefficient.
For these models, the decision variables are: $G_{local}^{inv}$, the fraction of a player's non-visible neighbors choosing $green$; $G_{local}^{vis}$, the fraction of a player's visible neighbors choosing $green$; $R_{local}^{inv}$, the fraction of a player's non-visible neighbors choosing $red$; and $R_{local}^{vis}$, the fraction of a player's visible neighbors choosing $red$. 
Note that $G_{local}^{inv} + R_{local}^{inv}$ and $G_{local}^{vis} + R_{local}^{vis}$ are not necessarily 1, since some of the neighbors may not have yet chosen a color.
All variables were normalized.
\begin{table}[h!]
	\centering
	\caption{Red picking model, P(red$ \arrowvert $ pick a color).}
	\label{table:parameters_color_choosing_2}       
	\begin{tabular}{ccccccc}
		\hline\noalign{\smallskip}
		Type & Visible neighbors & Intercept & $G_{local}^{inv}$ & $G_{local}^{vis}$ & $R_{local}^{inv}$ & $R_{local}^{vis}$\\
		\noalign{\smallskip}\hline\noalign{\smallskip}
		Regular & No & 0 & -4.863 &  & 5.032 &  \\ 
		& Yes &-0.066 & -2.855 & -2.022 & 3.453 & 1.733 \\ 
		Visible & No &0.109 & -4.411 &  & 4.202 &  \\ 
		& Yes &0.188 & -3.215 & -1.599 & 2.395 & 1.996 \\ 
		Adversarial & No &-0.023 & 0.817 &  & -0.649 &  \\ 
		& Yes &-0.286 & 0.172 & 0.732 & -0.204 &  \\ 
		\hline 
	\end{tabular}
\end{table}

The coefficients of the learned models are presented in Table~\ref{table:parameters_color_choosing_2}.
The results closely follow expectations here: the more neighbors (visible and not) are choosing \emph{red} as opposed to \emph{green}, the more likely the \emph{consensus} team player to choose \emph{red} as the initial color.
On the other hand, adversarial players tend to act in opposition to their neighbors, with \emph{red} prevalence in their local neighborhood generally leading them to choose \emph{green}.
The few noteworthy points, however, concern these adversarial nodes.
First, note that adversaries are much more influenced by visible nodes than non-visible neighbors (acting more strongly in opposition to these), whereas regular players tend to be less swayed by the behavior of visible neighbors as compared to others in their neighborhood.
Second, adversarial nodes act relatively unaggressively: the negative relationship between neighbor choices and their own initial color choice is relatively slight, in comparison with the magnitude of the positive relationships for the regular nodes.

\paragraph{\textbf{Timing of Color Change}}
Our last set of models determine the timing of a color change by a player.
More precisely, we again learn $l_1$-regularized logistic regression models which represent the probability that a player switches to the other color (either from \emph{red} to \emph{green}, or vice versa) at a given discrete time unit.
For these models, the decision variables are: 
$O_{local}^{inv}$, the fraction of a player's non-visible neighbors choosing the opposite color from the one chosen by the player; 
$O_{local}^{vis}$, the fraction of a player's visible neighbors choosing the opposite color from the one chosen by the player;
$C_{local}^{inv}$, the fraction of a player's non-visible neighbors choosing the same color as the player;
$C_{local}^{vis}$, the fraction of a player's visible neighbors choosing the same color as the player;
$N_{vis}$, the number of a player's neighbors who are visible; and
$N_{inv}$, the number of a player's neighbors that are non-visible players.

\begin{table}[h!]
	\caption{Color-changing model.}
	\label{table:parameters_color_changing} 
	\centering      
	\begin{tabular}{ccccccccc}
		\hline\noalign{\smallskip}
		Type & Visible neighbors & Intercept & $O_{local}^{inv}$ & $O_{local}^{vis}$ & $C_{local}^{inv}$ & $C_{local}^{vis}$ & $N_{inv}$ & $N_{vis}$\\
		\noalign{\smallskip}\hline\noalign{\smallskip}
		Regular & No & $-3.979$ & $2.6587$ & & $-0.330$ & & -0.01&\\
		& Yes & -3.79 & 1.1 & 1.484 & -0.874 & 0.095 & 0.004 & -0.034 \\ 
		Visible & No & 	-4.116 & 2.703 &  & -0.105 &  & -0.019 &  \\ 
		& Yes & 	-3.529 & 1.075 & 1.27 & -0.333 & -0.291 & -0.065 & 0.009 \\ 
		Adversarial & No &-2.799 & -1.131 &  & 1.191 &  & 0.007 &  \\ 
		& Yes & -2.723 & -0.599 & -0.372 & 0.948 & 0.306 & -0.002 & -0.198 \\ 
		\noalign{\smallskip}\hline
	\end{tabular}
\end{table}

The models' different coefficients are provided in Table~\ref{table:parameters_color_changing}.
Here the results are again somewhat intuitive: as we would expect, when the local color choices oppose that of a player, a regular player tends to switch, whereas the adversary tends to stay with their current color choice.
However, unlike their choice of the first color, here the adversaries less aggressively respond to visible node decisions as compared to those for their remaining neighbors.
On the other hand, they still tend to be somewhat less aggressive in acting against the neighborhood trends, as compared to \emph{consensus} players in their decisions to switch to be better aligned with these.
Our observation that adversarial nodes, while they follow a strategy one would intuitively expect, do so less aggressively than perhaps they could, is consistent with the behavioral observations we had previously made.
As we mentioned, this may be a part of an overall adversarial strategy to remain concealed, and thereby maximize impact.


\subsection{Analysis of Communication Behavior}

Given the importance of communication, we now consider the way in which the players are taking on different roles communicate.
First, we observe that, no matter what the role, the largest single class of messages attempt to stimulate coordination by naming a specific color (45\% of all messages).
Examples of this include messages that simply state a color (e.g., ``GREEN''), or suggest that everyone use a particular color (e.g., ``go for green'', or ``all green'').
We term all messages of this kind \emph{coordination} messages.
Another common form of communication is what we call \emph{information} messages (12\% of all messages), whereby players attempt to inform their network neighbors of their local state; an example of such a message would be ``5/5 red'', suggesting that 5 out of 5 neighbors of the node are choosing \emph{red}, or ``3r2g'', which communicates that 3 of the node's neighbors are choosing \emph{red} and two are choosing \emph{green}.

On average, a typical player sends quite a few messages, although this number varies dramatically depending on the player's role.
For example, a regular (non-visible) member of the \emph{consensus} team sends on average 20.6 messages each game (recall that games reaching consensus terminate immediately, so this implies that they communicate at least once every 3 seconds).
On the other hand, a visible node sends only 5.6 messages per game, and an adversarial node only 3.9 messages.
The fact that adversarial nodes make such limited use of the communication interface to prevent consensus is especially interesting---clearly, players taking on the role of adversaries are relatively unaggressive in this role.
In any case, when they do communicate, what do they write?

One thing we observe is that adversaries send considerably more coordination and information messages than \emph{consensus} players: 53\% and 15\%, respectively.
Thus, while adversarial nodes engage in considerably less communication, they appear to be more deliberate about it.
Next, we explore precisely how adversarial nodes use each of these two categories of messages towards their ends.

A natural strategy for an adversarial node in our setting is to send messages that are deliberately misleading.
We now explore the extent to which they do so for the two types of messages we identified above: \emph{coordination} and \emph{information} messages.
\begin{figure}[h!]
	\centering
	\includegraphics[width=0.5\linewidth]{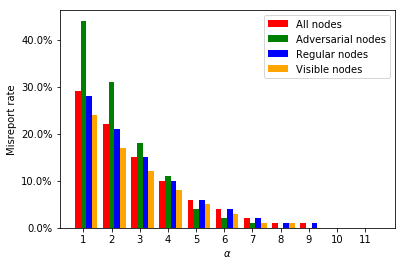}
	\caption{$ \alpha $-majority misreport rate.}
	\label{fig:misreport}
\end{figure}

First, consider the coordination messages.
Presumably, a misleading coordination message attempts to coordinate neighbors on a color which differs from that chosen by most of their neighbors.
However, such messages need not be \emph{deliberately} misleading.
With so many messages sent, there is bound to be a certain amount of noise in the nature of the messages.
Moreover, players may have a perception of what the likely consensus outcome is regardless of the particular current state of their local network (for example, when it appears that most of their neighbors are green most of the time, even if the majority of them happen to be choosing red at a particular point in time).
As a consequence, what is most important is the \emph{relative} rate with which such misleading messages are sent by adversaries, in comparison to other players.
These results are shown in Figure~\ref{fig:misreport}, as a function of the relative size of the majority.
Specifically, the $\alpha$ on the x-axis of the plot represents an $\alpha$-majority when at least $\alpha$ more players are choosing one color, and the coordination message is sent attempting to coordinate on another.
As we would expect, the fraction of such misreports (the misreport rate on the y-axis) drops quickly with increasing $\alpha$.
What is interesting is that, indeed, adversarial nodes are distinctly more misleading in this way than other nodes---a clear indication of such misleading messages being a part of a deliberate strategy.
However, no less interesting is the fact that once $\alpha > 3$, there is no longer a meaningful difference between adversarial players and others.
In other words, adversarial nodes attempt to be misleading, but only when it's not blatant.

Considering now information messages, we make a similar qualitative observation.
For such messages, we can quantify ``lying'' as incorrectly reporting local state.
Of course, we again must account for noise, in this case, erroneous reporting which is not deliberately a lie; thus, the focus is on the relative difference between reported and true state, in comparison with non-adversarial players.
We find that adversaries send information messages which are, indeed, more inaccurate on average than others.
Specifically, when the difference between reported and true state is normalized by the number of neighbors, adversaries are off by 0.5, in comparison with non-visible nodes, which are off by 0.3, and visible nodes, which are off by 0.4.
What we find, again, is that we see evidence that adversarial nodes deliberately lie about state, but such lies are rarely egregious.

One may wonder if the lack of aggressiveness on the part of adversaries is a sign of human cognitive limitation, or perhaps social consciousness (unwillingness to act in a way that causes harm to many others).
However, there is another natural explanation.
Recall that the identity of nodes (adversarial or not) is largely invisible.
An adversary who is overly aggressive may well reveal themselves as adversarial to all neighbors, who subsequently merely ignore them.
Thus, pulling punches may be a way to remain undetected, and may thereby be a sound strategy.

\section{Data-Driven Agent-Based Modeling and Analysis}
\label{S:ddabm}

The human subjects methodology is inherently limited in the number of experiments one can run and, consequently, the space of alternative configurations we can consider.
In the sequel, we engage in a further investigation of the problem of adversarial coordination using simulation experiments within an agent-based modeling framework. 
For this purpose, we make use of individual agent models developed in Section~\ref{S:individual}, and combine them into an agent-based model in which such artificial models are interacting on the exogenously specified networks.
While we found communication as an important factor in our analysis of the experiments, it is not clear how to model it in simulation, and we therefore focus on the setting with no communication and defer the issue of modeling communication to future work.

\subsection{Model Validation}


While statistical and face validity are essential steps in confirming that our individual behavior models are reasonable, we now add another dimension: validation in terms of \emph{aggregate outcomes} of agent-based simulations.
Specifically, we embed artificial agents into exogenous network topologies, as we had done with human subjects.
We then simulate identical environments as in our experiments, but now using artificial agents and in discrete time, for 60 iterations (since each time step in our models is equivalent to 1 second in the experiments).
Finally, we compare both qualitative trends, and quantitative outcomes, to those reported in the experimental results section above (Section~\ref{sec:ExpResults}).
Quantitatively, the agreement is reasonable, with the largest deviation between simulation outcomes and the experimental consensus rates within $0.14$.
\begin{figure}[h!]
	\centering
	\begin{tabular}{cc}
		\includegraphics[width=2in]{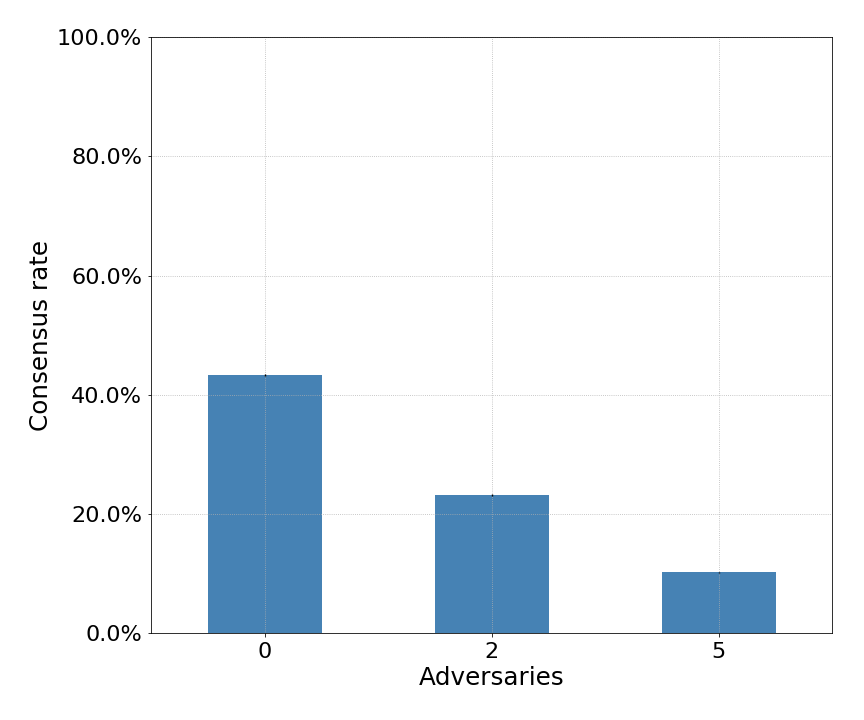} & 
		\includegraphics[width=2in]{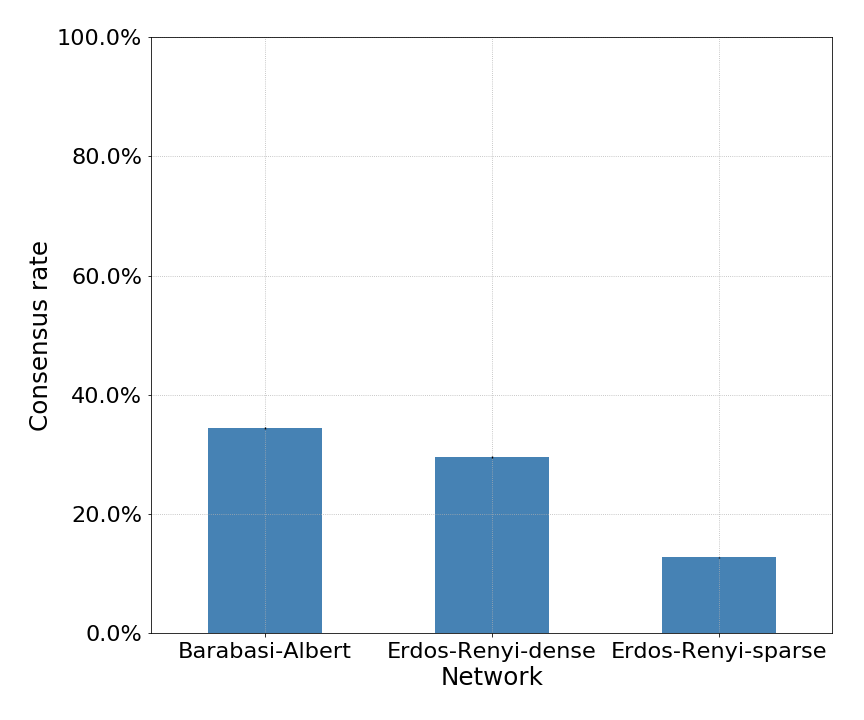}
	\end{tabular}
	\caption{Coordination ratios as a function of single variable.}
	\label{fig:results_over_1_par}
\end{figure}
The qualitative agreement is rather stronger, which we illustrate in Figure~\ref{fig:results_over_1_par}, which shows predicted consensus rates (using simulations) as a function of the number of adversaries (left plot) and network topology (right plot).
Comparing to corresponding results from the human subject experiments in Section \ref{sec:ExpResults}, we can observe broad qualitative agreement.
Note that agreement between simulated and experimental results we achieve for games at this scale (at least 20 players, with considerable interdependencies in behavior) compares quite favorably with similar efforts for devising artificial agents to model coordination in prior literature~\cite{Judd2010}.\footnote{\cite{Wunder13} is noteworthy as well.  However, they consider a public goods game, and aim to predict average contribution.  Predicting the probability of consensus using such data-driven agent-based simulations appears to be a more challenging problem.}


\subsection{Tuning Model Coefficients}

Human behavior captured in our experiments need not be rigid, and one could effect changes in behavior through a variety of techniques, such as presentation of choices, and increasing salience of some options through training~\cite{Thaler09}.
In this section, we consider the potential impact of small changes in behavior, as captured by the behavior model parameters, on the ability of people to successfully coordinate.
We focus on the models we learned to predict when players change their color.

Recall that for each non-adversarial player we have two models: the first when a player has at least one visible neighbor, and the second when they do not.
Since we have two types of non-adversarial actors (visible and non-visible nodes), we tune coefficients of four associated models, with the objective of maximizing consensus rate, with the constraint that the $l_1$ norm of the modification does not exceed an exogenously specified $\epsilon$.
We approximately solve this problem using \textit{Coordinate Greedy (CG)} local search, which iteratively chooses a parameter to optimize, and attempts to find the best improvement of this parameter.
To abide by the $l_1$ norm constraint, we subsequently project the result into the feasible space.

{
	We present the results of simulated consensus rates after parameters have been tuned in Figure \ref{fig:L1_norm_ret}, where
	the red dashed lines represent the coordination ratio when simulating with the original $\mathbf{w}$. 
	\begin{figure}[hbtp]
		\centering
		\begin{tabular}{cc}
			\includegraphics[width=0.45\linewidth]{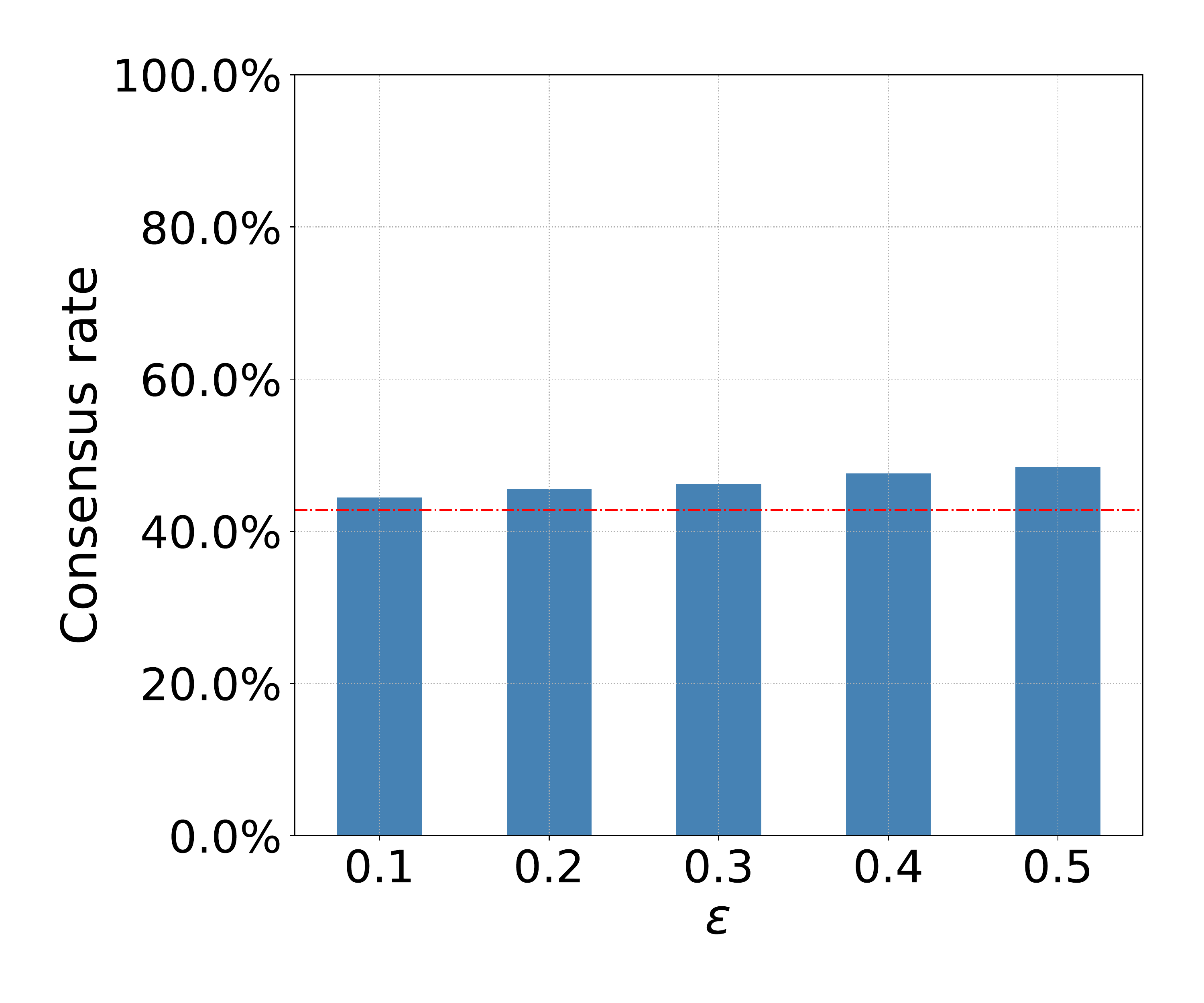} & \includegraphics[width=0.45\linewidth]{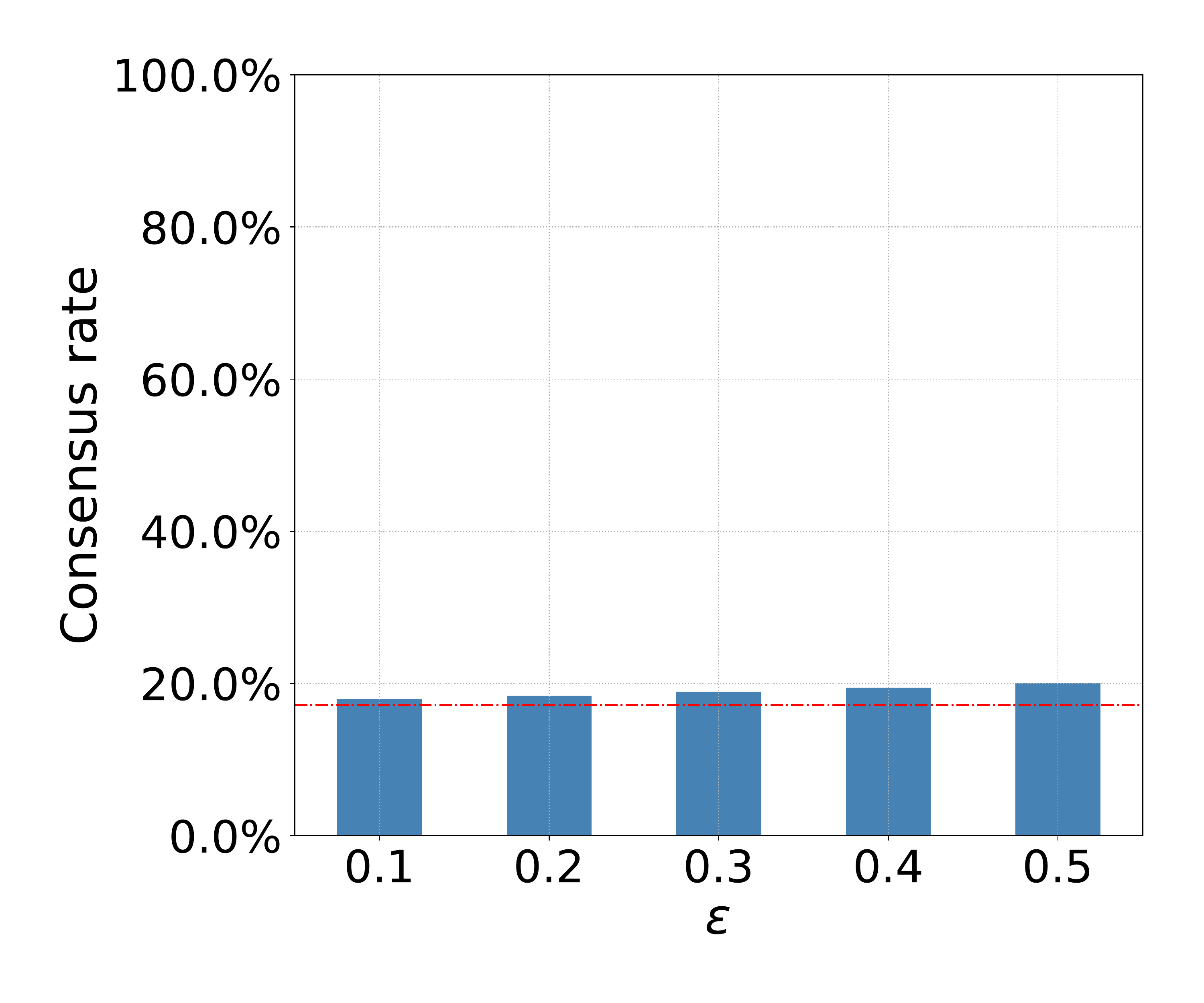}
		\end{tabular}
		\caption{$L_1$ norm constraint. Left: No adversaries setting. Right: With adversaries setting.}
		\label{fig:L1_norm_ret}
	\end{figure}
	Overall, even for relatively large $\epsilon$, the impact is surprisingly small: it appears that incremental changes in behavior of individuals has little impact on ability to successfully coordinate.
	This observation is especially clear in the adversarial setting.
}

\subsection{Placement of Trusted and Adversarial Players}

In our experiments, we randomly assigned trusted and adversarial players to nodes within the network.
We now explore the alternative possibility where the assignment of these is more deliberate.
To study the problem systematically, we consider the decision of where to place trusted (visible) and adversarial nodes in sequence, akin to a Stackelberg game: first, we choose placement for visible nodes, and second, given this placement, we choose placement for adversaries among the remaining nodes.
Unlike a Stackelberg game, however, we consider two approaches for choosing placement for visible and trusted nodes: random (as in our experiments), and by choosing a set of nodes maximizing the number of \emph{unique} neighbors (which we call \emph{optimal}).

\begin{figure}[h!]
	\centering
	\begin{tabular}{cc}
		\includegraphics[width=0.45\linewidth]{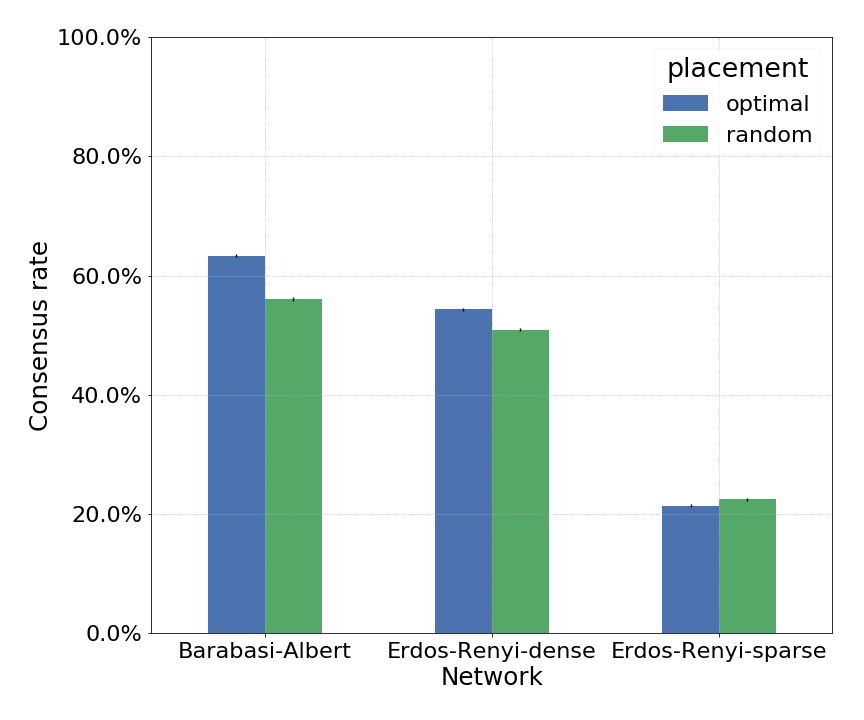} & 
		\includegraphics[width=0.45\linewidth]{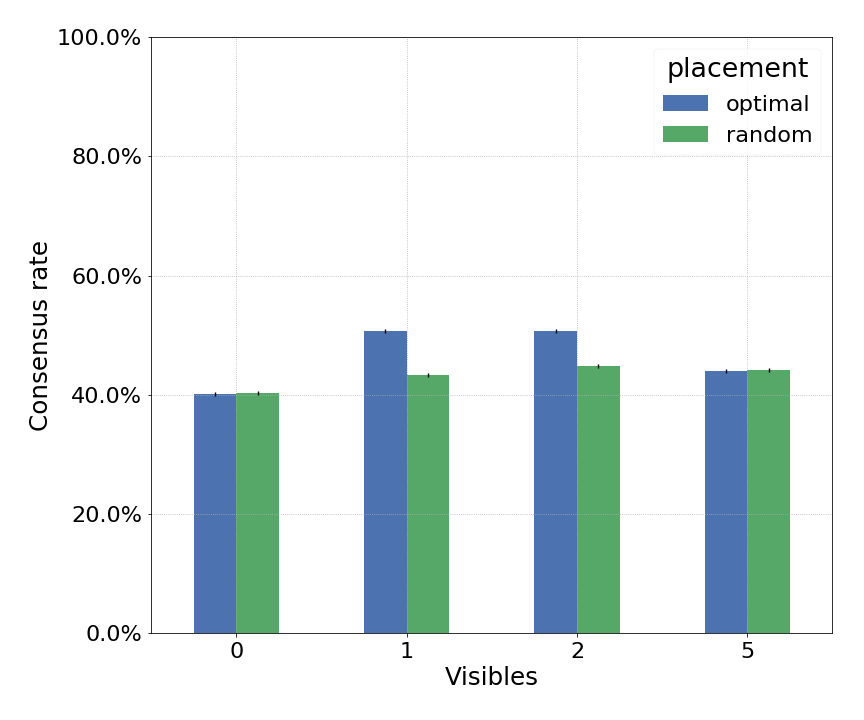}
	\end{tabular}
	\caption{Consensus rate as a function of placement of visible nodes when no adversaries are present. Left: for different network topologies. Right: number of visible nodes.}
	\label{fig:noAdv_placement}
\end{figure}
We first consider settings with no adversaries, and explore the impact of having an optimal placement of visible nodes, as compared with random placement.
The results are presented in Figure~\ref{fig:noAdv_placement}, for different network topologies (left), and different number of visible nodes (right).
First, we can see that the value of optimal placement is most pronounced in BA networks.
This is intuitive: such networks have few high-degree nodes, and making these visible can facilitate coordination.
Second, we can see an interesting phenomenon as we increase the number of visible players to 5, we can actually observe a \emph{decrease} in consensus rates, in the case when we greedily assign these to nodes.
We conjecture that the primary reason is that with 5 visible nodes, there is now a significant chance that some of them are not connected to each other, which can actually lead to a miscoordination of visible nodes themselves fail to agree on a color.
Since optimal placement of such nodes in the network ensures that visible nodes reach many others, miscoordination among visible nodes can thereby lead to miscoordination among a large number of others on the network.

\begin{figure*}[h!]
	\centering
	\begin{tabular}{ccc}
		\includegraphics[width=0.3\linewidth]{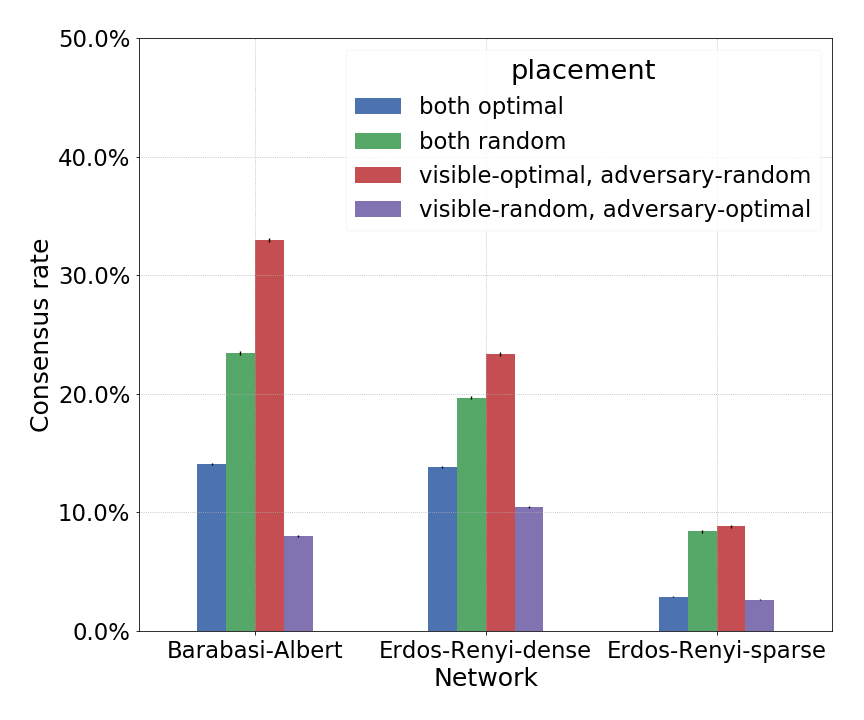} & 
		\includegraphics[width=0.3\linewidth]{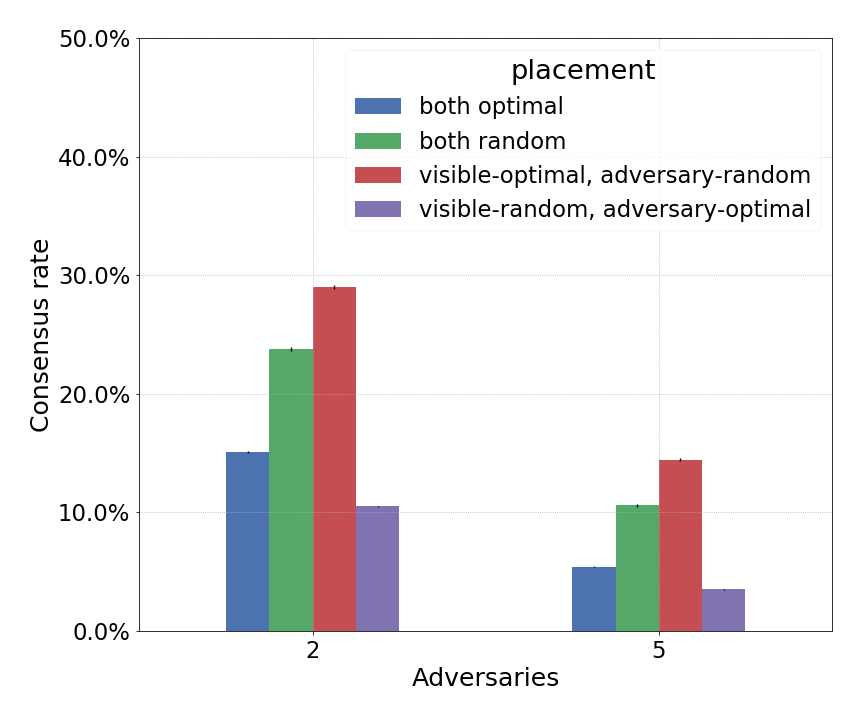} &
		\includegraphics[width=0.3\linewidth]{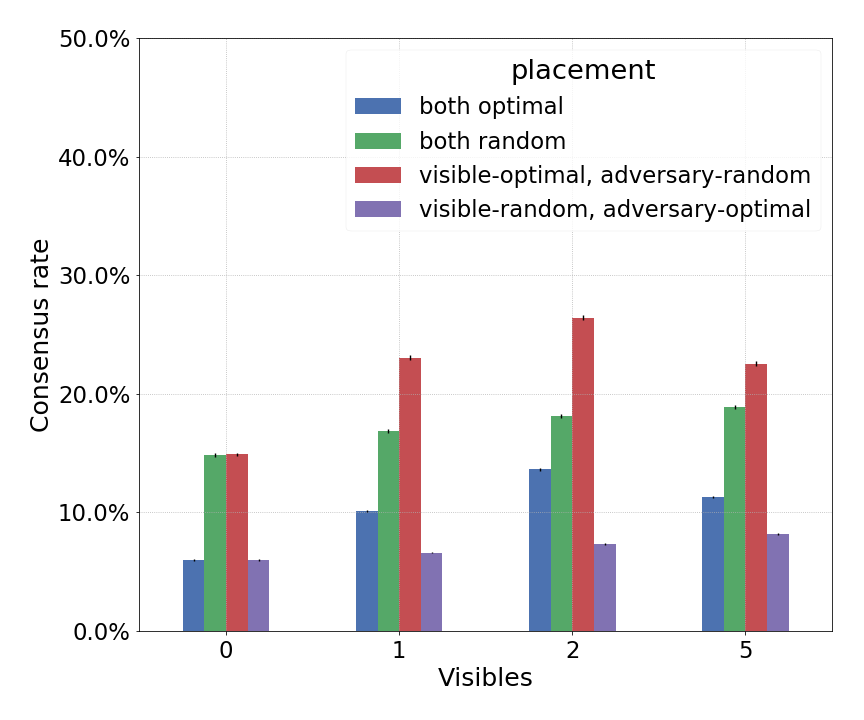} 
	\end{tabular}
	\caption{Consensus rate for different strategies of placing visible and adversarial nodes, as a function of: (Left) network topologies; (Middle) the number of adversaries; and (Right) the number of visible nodes.}
	\label{fig:withAdv_placement}
\end{figure*}
Figure~\ref{fig:withAdv_placement} presents the results of considering the two placement strategies (random and optimal) for visible and adversarial nodes.
From this figure we can make several noteworthy observations:
\begin{enumerate}
	\item {\it Adversarial players are highly effective with optimal placement}: consider blue and purple (first and last) bars in the plots, which correspond to adversaries placed greedily.  In both cases, consensus rates are quite small, for all network topologies, and even with only 2 adversaries.  This is especially surprising when we also consider the optimal placement of visible nodes, \emph{which are placed before adversaries}, and can thereby ensure that networks remain connected even after adversarial nodes are added.  While optimal placement of visible nodes clearly helps, the impact is smaller than we would have expected.  This shows that the additional behavioral patterns followed by adversarial nodes are, indeed, quite effective in hampering the ability of the \emph{consensus} team to coordinate successfully.
	\item {\it Greedily placing visible nodes helps}: consider the red bars (tallest in all plots), which correspond to the optimal placement of visible nodes, followed by random placement of adversaries.  In this situation, we can observe a clear value of visible nodes, particularly for the scale-free (BA) topology.  On the other hand, we can see that having 2 visible nodes is actually better than 5, again, due to the increased potential for miscoordination among visible nodes themselves in the latter case.
\end{enumerate}

\subsection{Impact of Network Topology}
In this section, we systematically explore the impact of network topological characteristics on consensus rate. 
For BA networks, we consider two parameters: $m$, the number of connections we add to each node entering the network, which controls density, and $\gamma$, where the probability of connecting to a node with degree $d$ is proportional to $d^\gamma$, which determines how heavy the tail of the distribution is.
For ER networks, we vary the probability $p$ that a pair of nodes is connected, which is also directly related to density.
Finally, we consider small-world networks~\cite{watts1998collective}, and vary the clustering coefficient.

\begin{figure*}[h!]
	\centering
	\begin{tabular}{ccc}
		\includegraphics[width=0.3\linewidth]{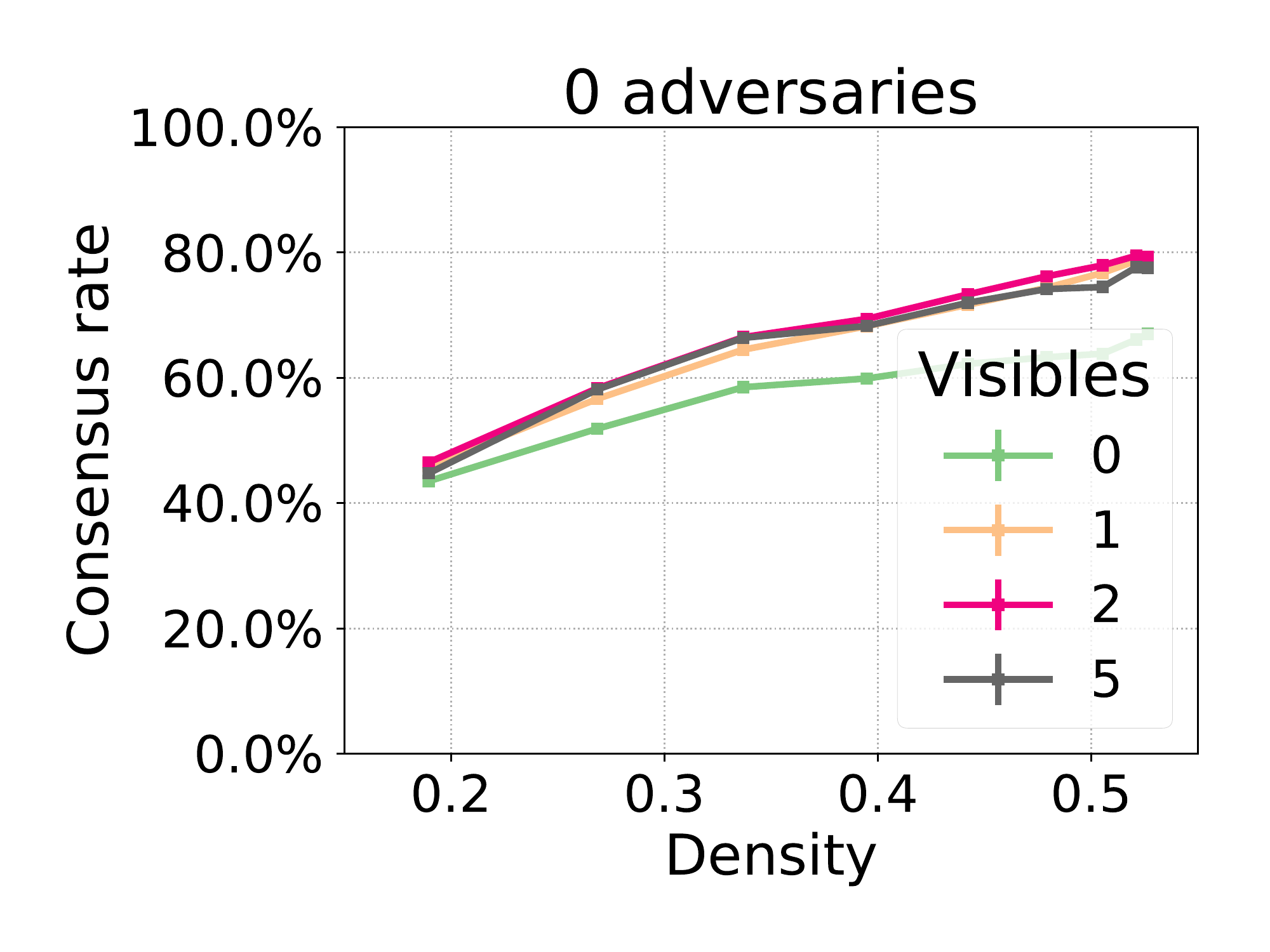} &
		\includegraphics[width=0.3\linewidth]{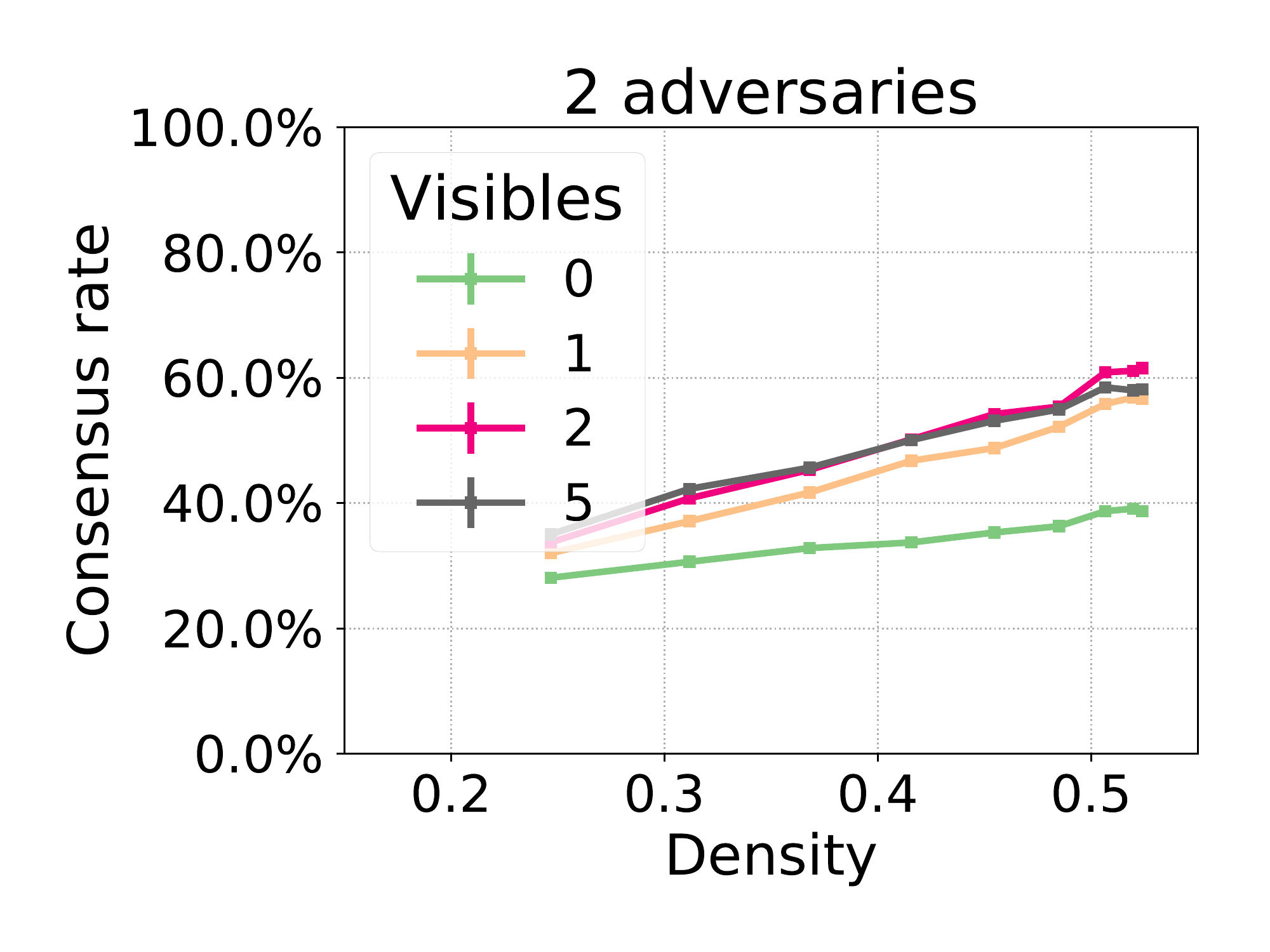} &
		\includegraphics[width=0.3\linewidth]{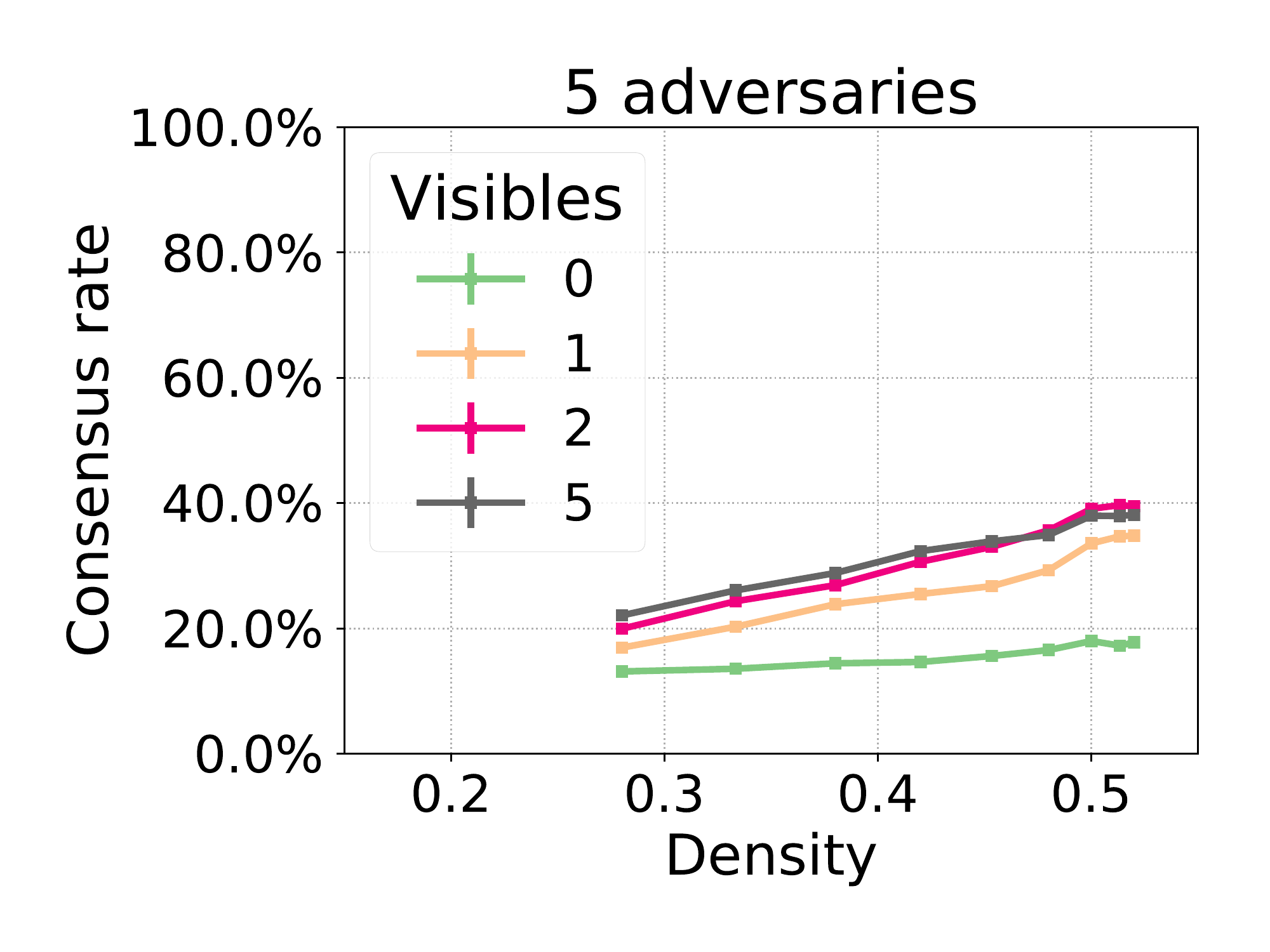}
	\end{tabular}
	\caption{Consensus rate in BA networks as a function of network density, broken down by the number of adversaries and the number of visible nodes. Left: 0 adversaries. Middle: 2 adversaries. Right: 5 adversaries.  
	}
	\label{fig:density}
\end{figure*}
\begin{figure*}[htbp]
	\centering
	\begin{tabular}{ccc}
		\includegraphics[width=0.3\linewidth]{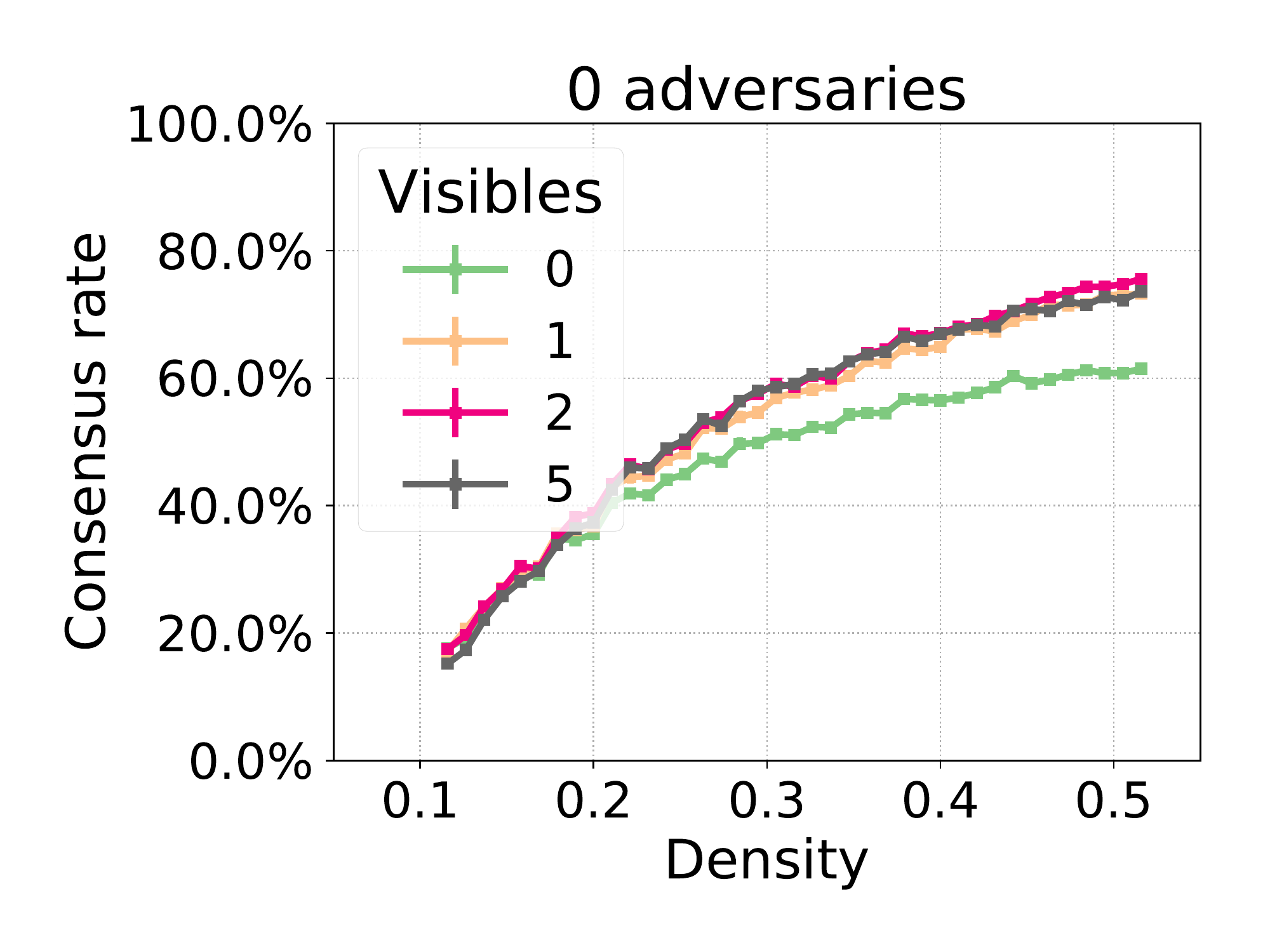} &
		\includegraphics[width=0.3\linewidth]{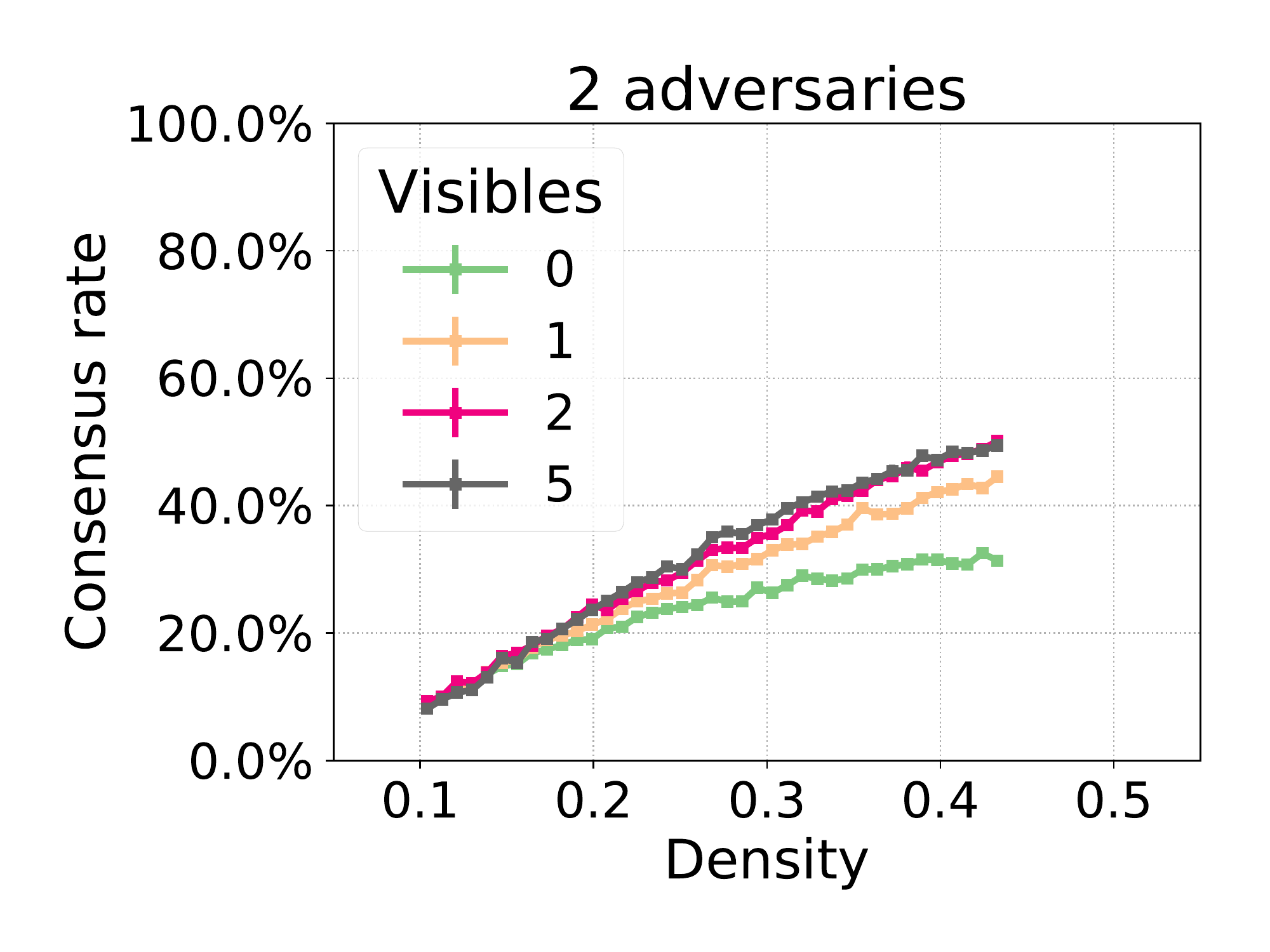} &
		\includegraphics[width=0.3\linewidth]{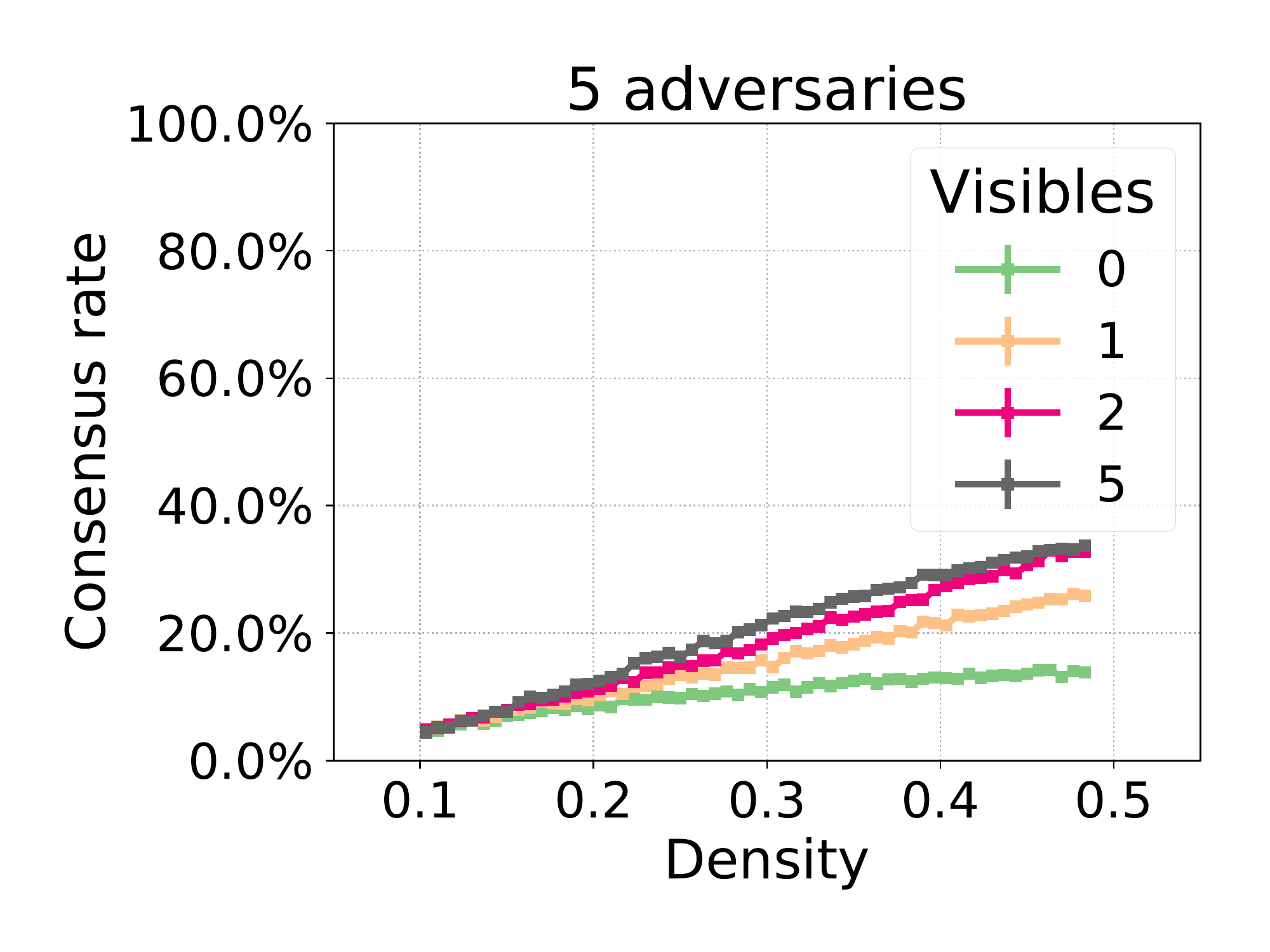} 
	\end{tabular}
	\caption{Consensus rate as a function of network density, broken down by the number of adversaries and the number of visible nodes. Left: 0 adversaries; Middle: 2 adversaries; Right: 5 adversaries.}
	\label{fig:ER_density_visible}
\end{figure*}

Figure~\ref{fig:density} shows the trends in consensus rates for different numbers of adversaries and visible nodes, as a function of network density.
We only show the results for the BA topology; there is little qualitative difference for ER topologies (Figure~\ref{fig:ER_density_visible}).
Overall, increased density tends to improve consensus rates, with and without adversaries.
More interestingly, the presence of visible nodes becomes more valuable with increased density as well, albeit 1 such node generally seems to suffice.

\begin{figure*}[htbp]
	\centering
	\begin{tabular}{ccc}
		\includegraphics[width=0.3\linewidth]{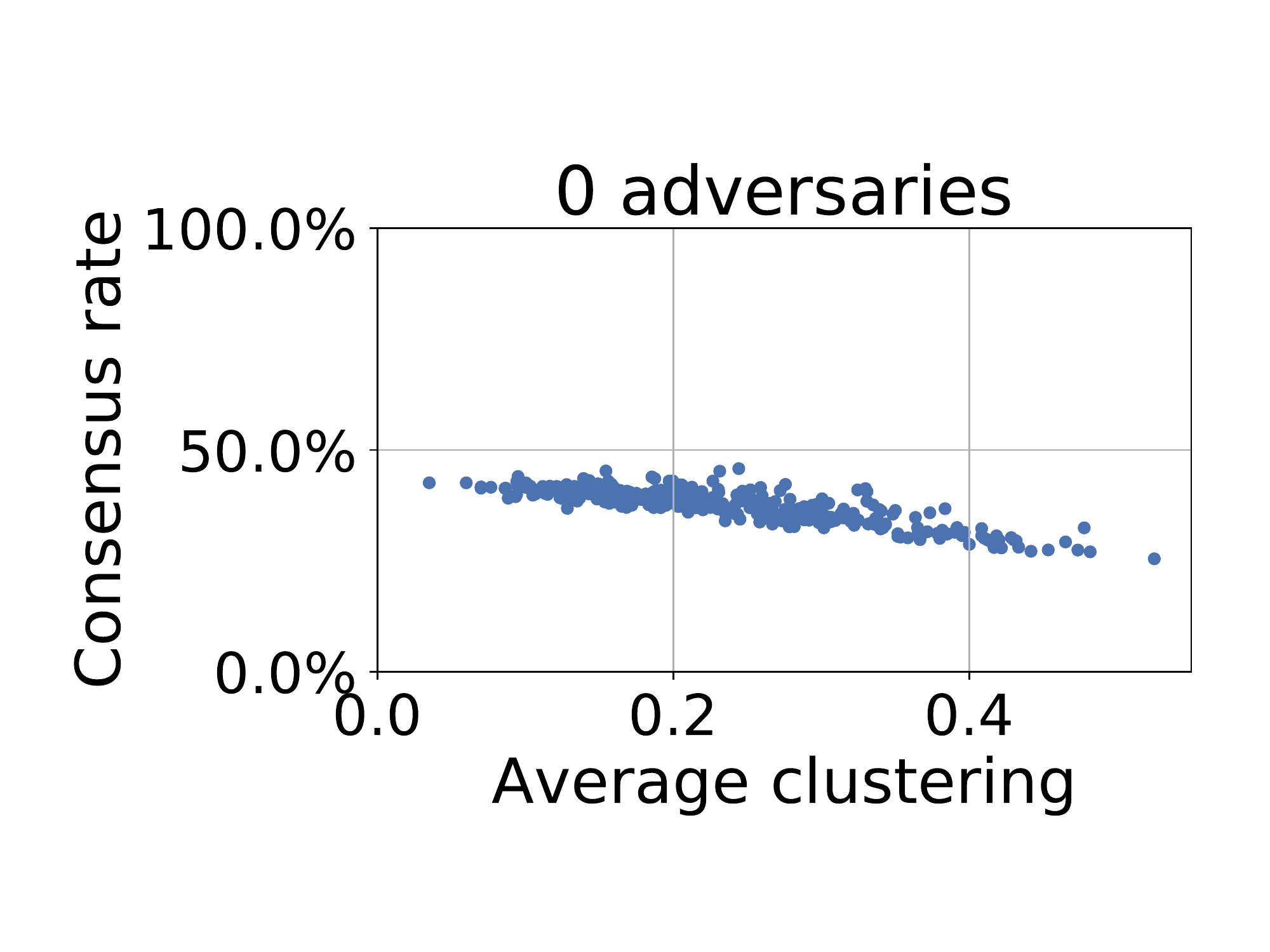} &
		\includegraphics[width=0.3\linewidth]{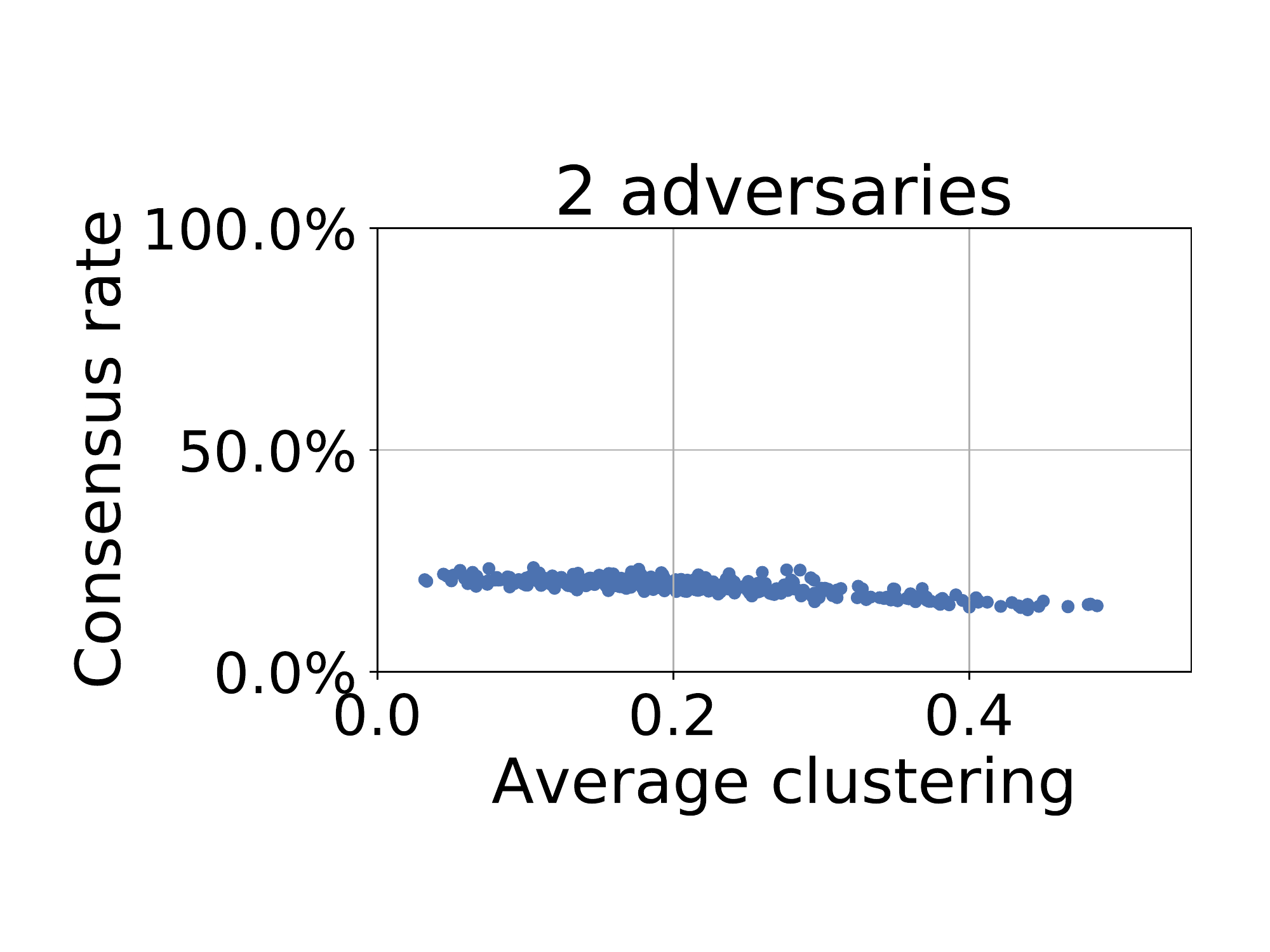} &
		\includegraphics[width=0.3\linewidth]{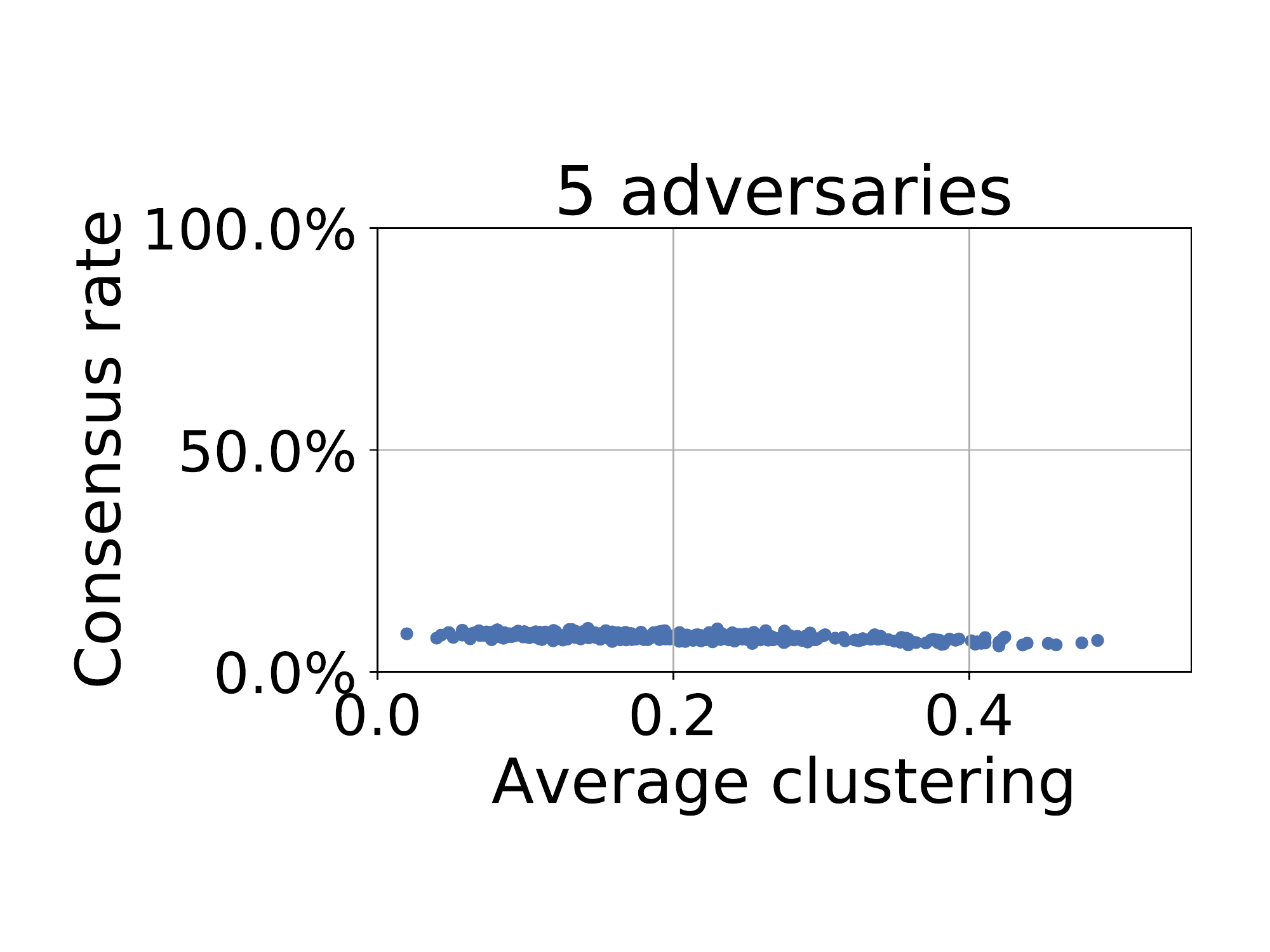} 
	\end{tabular}
	\caption{Consensus rate as a function of average clustering coefficient, broken down by the number of adversaries. Left: 0 adversaries. Middle: 2 adversaries. Right: 5 adversaries.}
	\label{fig:SmallWorld_avgClustering}
\end{figure*}
Figure~\ref{fig:SmallWorld_avgClustering} shows the impact of increasing the clustering coefficient (keeping density fixed).
Here, we see that the trend is that higher clustering tends to hurt coordination, a finding that echoes previously reported results~\cite{Kearns2009}.
However, the trend becomes flatter as we add adversaries.
We found that adding visible nodes, in this case, has no tangible impact on consensus rates.

\begin{figure*}[htbp]
	\centering
	\begin{tabular}{ccc}
		\includegraphics[width=0.3\linewidth]{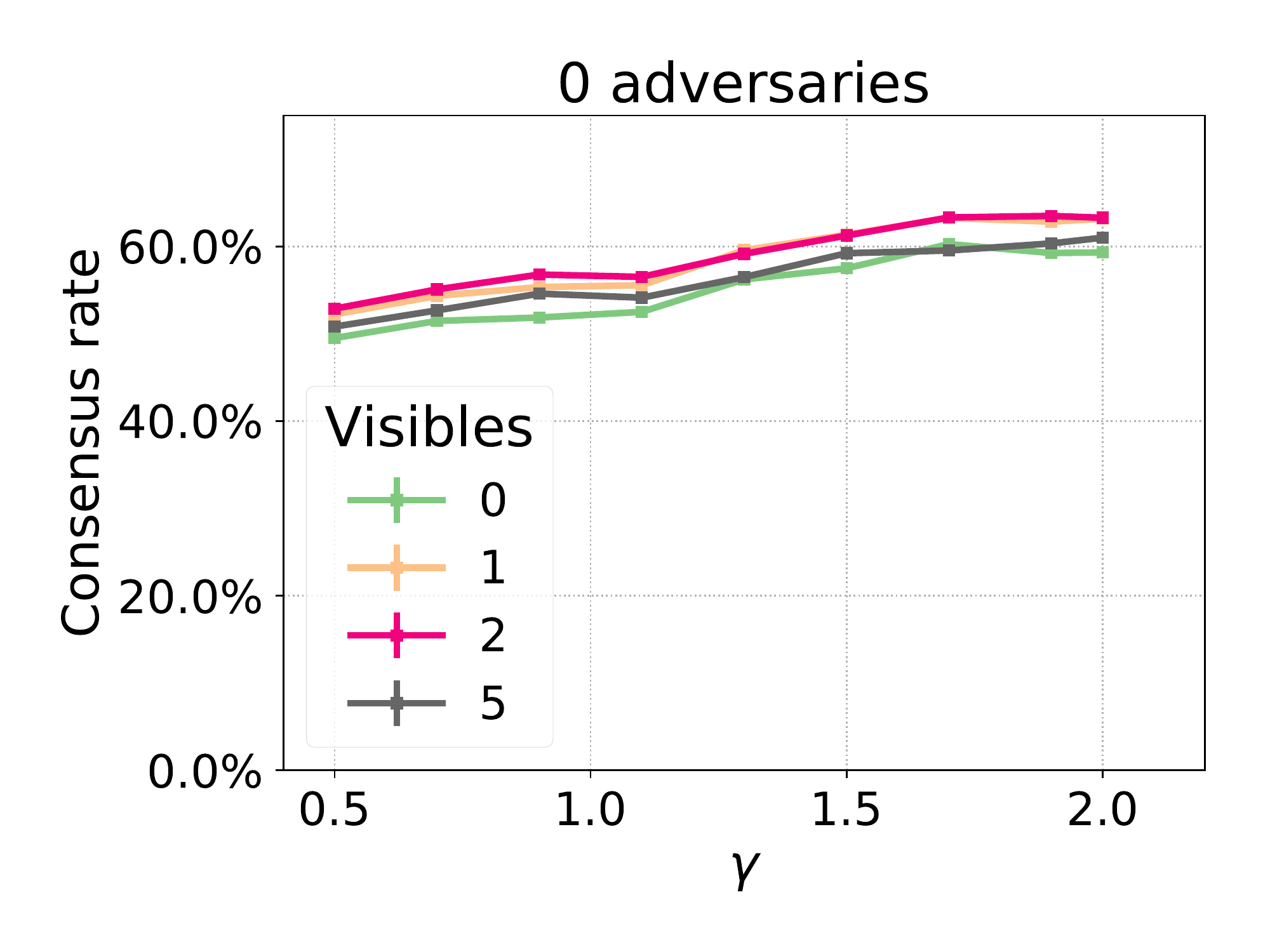} &
		\includegraphics[width=0.3\linewidth]{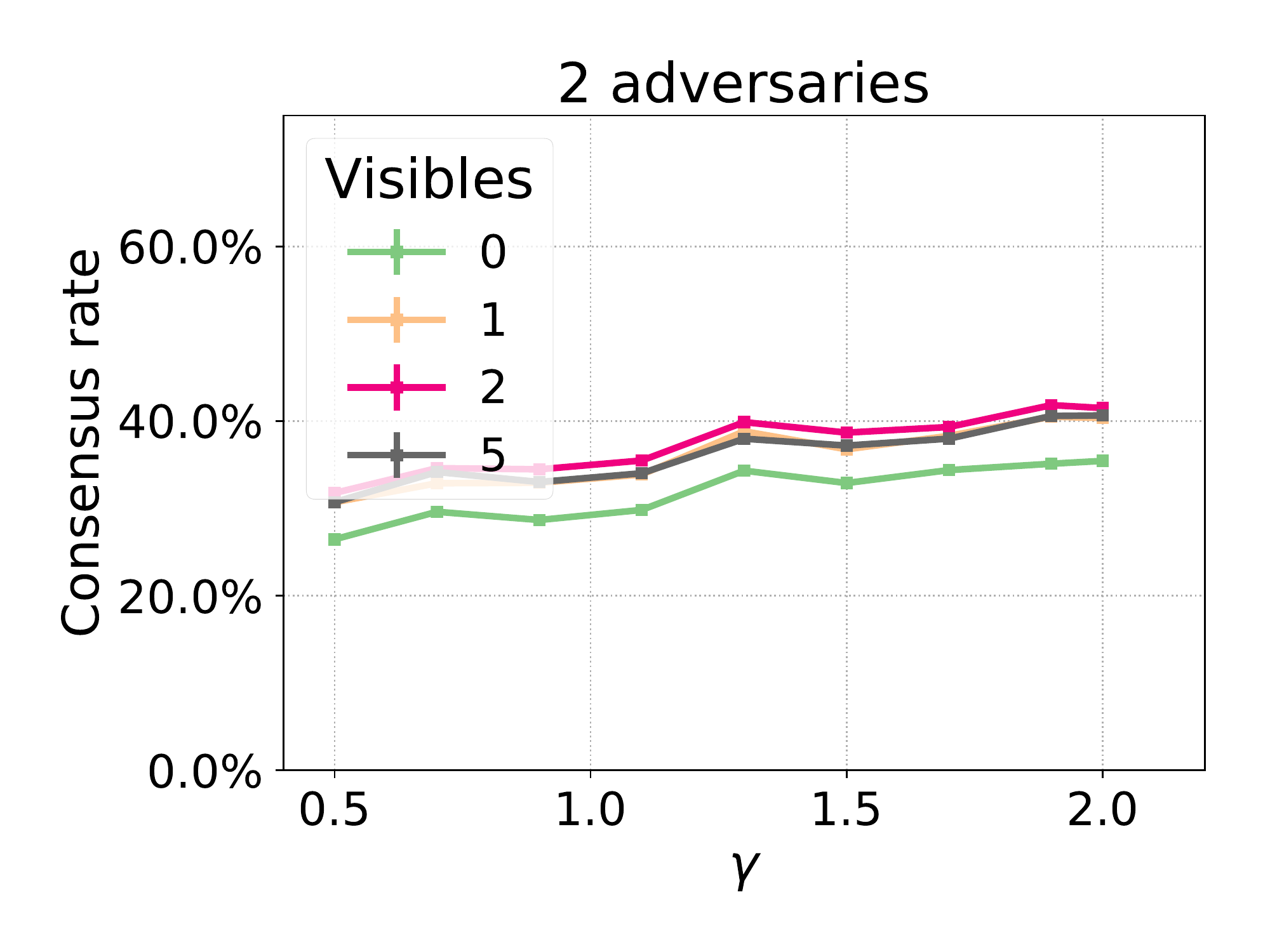} &
		\includegraphics[width=0.3\linewidth]{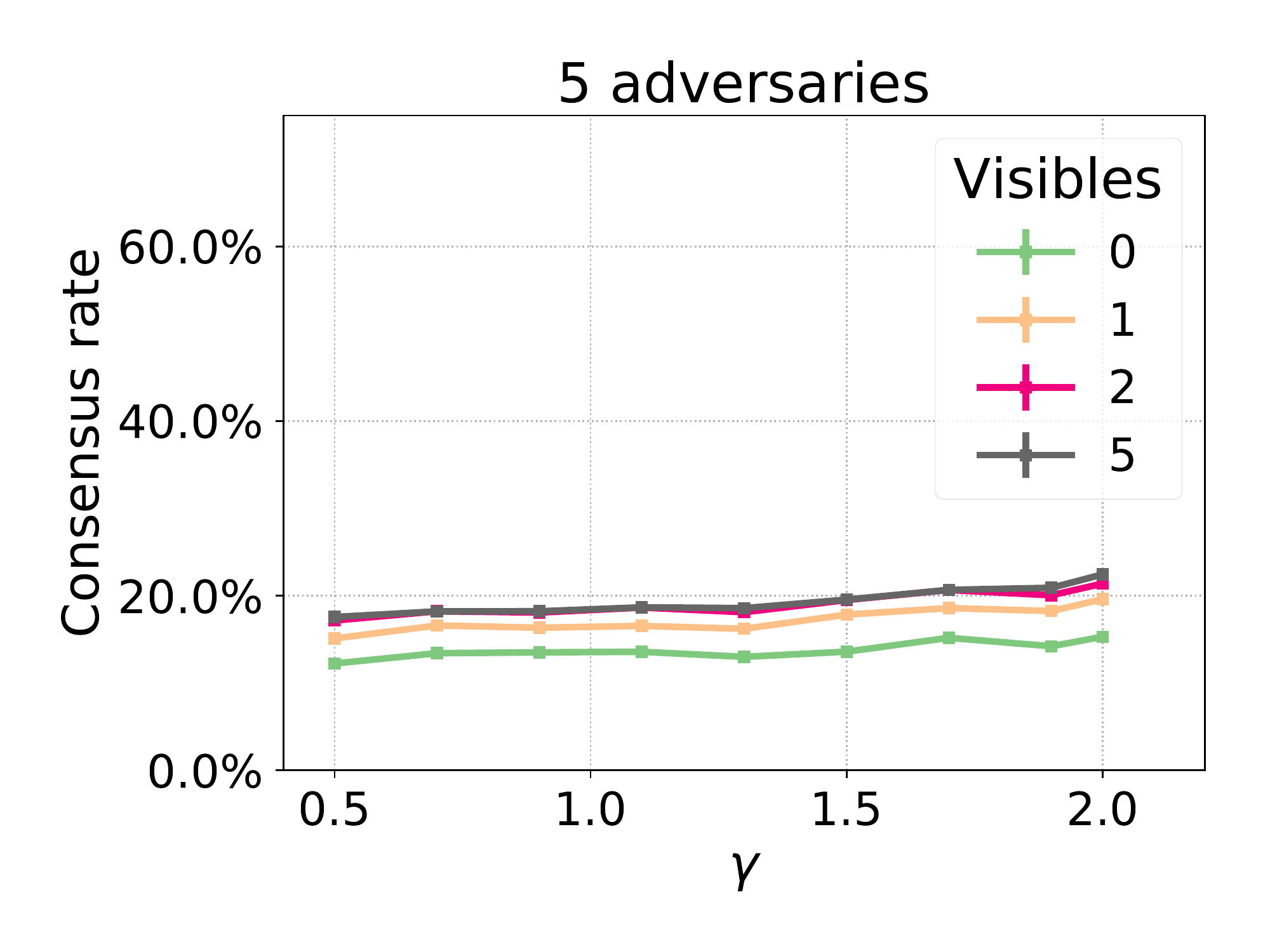} 
	\end{tabular}
	\caption{Consensus rate as a function of $\gamma$, broken down by the number of adversaries. Left: 0 adversaries. Middle: 2 adversaries. Right: 5 adversaries.}
	\label{fig:PA_varyingGamma}
\end{figure*}
Figure~\ref{fig:PA_varyingGamma} shows consensus rate as a function of $\gamma$ (higher implies greater disparity in degrees).
In general, degree distributions with a heavier tail yield higher consensus rates, as long as there are only a few adversaries; the relationship becomes essentially flat with 5 adversaries.
The reason is that heavy-tail distributions have fewer central actors with more neighbors, and as long as these are not adversarial, they can considerably facilitate consensus.
Since we assign adversarial nodes randomly in these experiments, it is unlikely that any such high-degree nodes are adversarial if there are only 2 adversaries, but this becomes far more likely with 5 adversaries.

\section{Conclusion}

We consider the problem of adversarial consensus on social networks both using human subjects and simulation-based methodologies.
The overall goal of the subjects is to reach global consensus on a particular color, despite adversarial nodes who attempt to prevent consensus.
We find that adversarial players are effective at significantly reducing consensus rates.
While the ability to communicate can significantly improve coordination success, particularly as we increase the number of adversaries, embedding trusted nodes within the network is of limited value.
We observe several strategies used by adversarial players to subvert coordination, including choosing a color which opposes local majority and communicating misleading information and instructions to their neighbors.
However, we also note that these malicious activities are used in a somewhat subdued manner, suggesting perhaps an attempt of adversarial players to remain covert.
Extensive simulations using an agent-based model created based on experimental data additionally show that the importance of trusted nodes does increase as we increase network density, but small changes to parameters of the behavior models yield limited impact on consensus rates.

\bibliographystyle{ACM-Reference-Format}  
\bibliography{Coordination,refs}  


\begin{thebibliography}{00}


\ifx \showCODEN    \undefined \def \showCODEN     #1{\unskip}     \fi
\ifx \showDOI      \undefined \def \showDOI       #1{{\tt DOI:}\penalty0{#1}\ }
  \fi
\ifx \showISBNx    \undefined \def \showISBNx     #1{\unskip}     \fi
\ifx \showISBNxiii \undefined \def \showISBNxiii  #1{\unskip}     \fi
\ifx \showISSN     \undefined \def \showISSN      #1{\unskip}     \fi
\ifx \showLCCN     \undefined \def \showLCCN      #1{\unskip}     \fi
\ifx \shownote     \undefined \def \shownote      #1{#1}          \fi
\ifx \showarticletitle \undefined \def \showarticletitle #1{#1}   \fi
\ifx \showURL      \undefined \def \showURL       #1{#1}          \fi
\providecommand\bibfield[2]{#2}
\providecommand\bibinfo[2]{#2}
\providecommand\natexlab[1]{#1}
\providecommand\showeprint[2][]{arXiv:#2}

\bibitem[\protect\citeauthoryear{Abbas, Vorobeychik, and Koutsoukos}{Abbas
  et~al\mbox{.}}{2014}]%
        {Abbas14}
\bibfield{author}{\bibinfo{person}{W. Abbas}, \bibinfo{person}{Y. Vorobeychik},
  {and} \bibinfo{person}{X. Koutsoukos}.} \bibinfo{year}{2014}\natexlab{}.
\newblock \showarticletitle{Resilient consensus protocol in the presence of
  trusted nodes}. In \bibinfo{booktitle}{{\em 2014 7th International Symposium
  on Resilient Control Systems (ISRCS)}}. \bibinfo{pages}{1--7}.
\newblock
\showDOI{%
\url{http://dx.doi.org/10.1109/ISRCS.2014.6900100}}


\bibitem[\protect\citeauthoryear{Albert, Jeong, and Barabasi}{Albert
  et~al\mbox{.}}{2000}]%
        {Albert00}
\bibfield{author}{\bibinfo{person}{R. Albert}, \bibinfo{person}{H. Jeong},
  {and} \bibinfo{person}{A.-L. Barabasi}.} \bibinfo{year}{2000}\natexlab{}.
\newblock \showarticletitle{Error and attack tolerance of complex networks}.
\newblock \bibinfo{journal}{{\em Nature\/}}  \bibinfo{volume}{406}
  (\bibinfo{year}{2000}), \bibinfo{pages}{378--482}.
\newblock


\bibitem[\protect\citeauthoryear{Barabasi and Albert}{Barabasi and
  Albert}{1999}]%
        {Barabasi99}
\bibfield{author}{\bibinfo{person}{Albert-Laszlo Barabasi} {and}
  \bibinfo{person}{Reka Albert}.} \bibinfo{year}{1999}\natexlab{}.
\newblock \showarticletitle{Emergence of Scaling in Random Networks}.
\newblock \bibinfo{journal}{{\em Science\/}} \bibinfo{volume}{286},
  \bibinfo{number}{5439} (\bibinfo{year}{1999}), \bibinfo{pages}{509--512}.
\newblock


\bibitem[\protect\citeauthoryear{Chakraborty, Judd, Kearns, and
  Tan}{Chakraborty et~al\mbox{.}}{2010}]%
        {Chakraborty10}
\bibfield{author}{\bibinfo{person}{Tanmoy Chakraborty},
  \bibinfo{person}{Stephen Judd}, \bibinfo{person}{Michael Kearns}, {and}
  \bibinfo{person}{Jinsong Tan}.} \bibinfo{year}{2010}\natexlab{}.
\newblock \showarticletitle{A Behavioral Study of Bargaining in Social
  Networks}. In \bibinfo{booktitle}{{\em Proceedings of the 11th ACM Conference
  on Electronic Commerce}}. \bibinfo{pages}{243--252}.
\newblock


\bibitem[\protect\citeauthoryear{Cooper, DeJong, Forsythe, and Ross}{Cooper
  et~al\mbox{.}}{1992}]%
        {Cooper1992}
\bibfield{author}{\bibinfo{person}{Russell Cooper}, \bibinfo{person}{Douglas~V.
  DeJong}, \bibinfo{person}{Robert Forsythe}, {and} \bibinfo{person}{Thomas~W.
  Ross}.} \bibinfo{year}{1992}\natexlab{}.
\newblock \showarticletitle{{Communication in coordination games}}.
\newblock \bibinfo{journal}{{\em Quarterly Journal of Economics\/}}
  \bibinfo{volume}{107}, \bibinfo{number}{2} (\bibinfo{year}{1992}),
  \bibinfo{pages}{739--771}.
\newblock


\bibitem[\protect\citeauthoryear{Demichelis and Weibull}{Demichelis and
  Weibull}{2008}]%
        {Demichelis2008}
\bibfield{author}{\bibinfo{person}{Stefano Demichelis} {and}
  \bibinfo{person}{Jorgen~W. Weibull}.} \bibinfo{year}{2008}\natexlab{}.
\newblock \showarticletitle{{Language, meaning, and games: A model of
  communication, coordination, and evolution}}.
\newblock \bibinfo{journal}{{\em American Economic Review\/}}
  \bibinfo{volume}{98}, \bibinfo{number}{4} (\bibinfo{year}{2008}),
  \bibinfo{pages}{1292--1311}.
\newblock


\bibitem[\protect\citeauthoryear{Ellingsen and Ostling}{Ellingsen and
  Ostling}{2010}]%
        {Ellingsen2010}
\bibfield{author}{\bibinfo{person}{Tore Ellingsen} {and}
  \bibinfo{person}{Robert Ostling}.} \bibinfo{year}{2010}\natexlab{}.
\newblock \showarticletitle{{When does communication improve coordination?}}
\newblock \bibinfo{journal}{{\em American Economic Review\/}}
  \bibinfo{volume}{100} (\bibinfo{year}{2010}), \bibinfo{pages}{1695--1724}.
\newblock


\bibitem[\protect\citeauthoryear{Erdos and Renyi}{Erdos and Renyi}{1960}]%
        {Erdos60}
\bibfield{author}{\bibinfo{person}{P. Erdos} {and} \bibinfo{person}{A. Renyi}.}
  \bibinfo{year}{1960}\natexlab{}.
\newblock \showarticletitle{On the evolution of random graphs}.
\newblock \bibinfo{journal}{{\em Publ. Math. Inst. Hung. Acad. Sci\/}}
  \bibinfo{volume}{5}, \bibinfo{number}{17} (\bibinfo{year}{1960}),
  \bibinfo{pages}{17--61}.
\newblock


\bibitem[\protect\citeauthoryear{Farrell}{Farrell}{1987}]%
        {Farrell87}
\bibfield{author}{\bibinfo{person}{Joseph Farrell}.}
  \bibinfo{year}{1987}\natexlab{}.
\newblock \showarticletitle{Cheap talk, coordination, and entry}.
\newblock \bibinfo{journal}{{\em RAND Journal of Economics\/}}
  \bibinfo{volume}{18}, \bibinfo{number}{1} (\bibinfo{year}{1987}),
  \bibinfo{pages}{34--39}.
\newblock


\bibitem[\protect\citeauthoryear{Farrell}{Farrell}{1988}]%
        {Farrell88}
\bibfield{author}{\bibinfo{person}{Joseph Farrell}.}
  \bibinfo{year}{1988}\natexlab{}.
\newblock \showarticletitle{Communication, coordination and {Nash}
  equilibrium}.
\newblock \bibinfo{journal}{{\em Economic Letters\/}}  \bibinfo{volume}{27}
  (\bibinfo{year}{1988}), \bibinfo{pages}{209--214}.
\newblock


\bibitem[\protect\citeauthoryear{Gracia-L{\'a}zaro, Ferrer, Ruiz, Taranc{\'o}n,
  Cuesta, S{\'a}nchez, and Moreno}{Gracia-L{\'a}zaro et~al\mbox{.}}{2012}]%
        {Gracia12}
\bibfield{author}{\bibinfo{person}{Carlos Gracia-L{\'a}zaro},
  \bibinfo{person}{Alfredo Ferrer}, \bibinfo{person}{Gonzalo Ruiz},
  \bibinfo{person}{Alfonso Taranc{\'o}n}, \bibinfo{person}{Jos{\'e}~A. Cuesta},
  \bibinfo{person}{Angel S{\'a}nchez}, {and} \bibinfo{person}{Yamir Moreno}.}
  \bibinfo{year}{2012}\natexlab{}.
\newblock \showarticletitle{Heterogeneous networks do not promote cooperation
  when humans play a Prisoner{\textquoteright}s Dilemma}.
\newblock \bibinfo{journal}{{\em Proceedings of the National Academy of
  Sciences\/}} \bibinfo{volume}{109}, \bibinfo{number}{32}
  (\bibinfo{year}{2012}), \bibinfo{pages}{12922--12926}.
\newblock


\bibitem[\protect\citeauthoryear{Hajaj, Hazon, and Sarne}{Hajaj
  et~al\mbox{.}}{2015}]%
        {hajaj2015improving}
\bibfield{author}{\bibinfo{person}{Chen Hajaj}, \bibinfo{person}{Noam Hazon},
  {and} \bibinfo{person}{David Sarne}.} \bibinfo{year}{2015}\natexlab{}.
\newblock \showarticletitle{Improving comparison shopping agents’ competence
  through selective price disclosure}.
\newblock \bibinfo{journal}{{\em Electronic Commerce Research and
  Applications\/}} \bibinfo{volume}{14}, \bibinfo{number}{6}
  (\bibinfo{year}{2015}), \bibinfo{pages}{563--581}.
\newblock


\bibitem[\protect\citeauthoryear{Hajaj, Hazon, and Sarne}{Hajaj
  et~al\mbox{.}}{2017}]%
        {hajaj2017enhancing}
\bibfield{author}{\bibinfo{person}{Chen Hajaj}, \bibinfo{person}{Noam Hazon},
  {and} \bibinfo{person}{David Sarne}.} \bibinfo{year}{2017}\natexlab{}.
\newblock \showarticletitle{Enhancing comparison shopping agents through
  ordering and gradual information disclosure}.
\newblock \bibinfo{journal}{{\em Autonomous Agents and Multi-Agent Systems\/}}
  \bibinfo{volume}{31}, \bibinfo{number}{3} (\bibinfo{year}{2017}),
  \bibinfo{pages}{696--714}.
\newblock


\bibitem[\protect\citeauthoryear{Judd, Kearns, and Vorobeychik}{Judd
  et~al\mbox{.}}{2010}]%
        {Judd2010}
\bibfield{author}{\bibinfo{person}{Stephen Judd}, \bibinfo{person}{Michael
  Kearns}, {and} \bibinfo{person}{Yevgeniy Vorobeychik}.}
  \bibinfo{year}{2010}\natexlab{}.
\newblock \showarticletitle{{Behavioral dynamics and influence in networked
  coloring and consensus}}.
\newblock \bibinfo{journal}{{\em Proceedings of the National Academy of
  Sciences\/}} \bibinfo{volume}{107}, \bibinfo{number}{34}
  (\bibinfo{year}{2010}), \bibinfo{pages}{14978--14982}.
\newblock


\bibitem[\protect\citeauthoryear{Kearns}{Kearns}{2012}]%
        {Kearns2012}
\bibfield{author}{\bibinfo{person}{Michael Kearns}.}
  \bibinfo{year}{2012}\natexlab{}.
\newblock \showarticletitle{{Experiments in social computation}}.
\newblock \bibinfo{journal}{{\it Commun. ACM}} \bibinfo{volume}{55},
  \bibinfo{number}{10} (\bibinfo{year}{2012}), \bibinfo{pages}{56--67}.
\newblock


\bibitem[\protect\citeauthoryear{Kearns, Judd, Tan, and Wortman}{Kearns
  et~al\mbox{.}}{2009}]%
        {Kearns2009}
\bibfield{author}{\bibinfo{person}{Michael Kearns}, \bibinfo{person}{Stephen
  Judd}, \bibinfo{person}{Jinsong Tan}, {and} \bibinfo{person}{Jennifer
  Wortman}.} \bibinfo{year}{2009}\natexlab{}.
\newblock \showarticletitle{{Behavioral experiments in biased voting in
  networks}}.
\newblock \bibinfo{journal}{{\em Proceedings of the National Academy of
  Sciences\/}} \bibinfo{volume}{106}, \bibinfo{number}{5}
  (\bibinfo{year}{2009}), \bibinfo{pages}{1347--1352}.
\newblock


\bibitem[\protect\citeauthoryear{Kearns, Judd, and Vorobeychik}{Kearns
  et~al\mbox{.}}{2012}]%
        {kearns2012behavioral}
\bibfield{author}{\bibinfo{person}{Michael Kearns}, \bibinfo{person}{Stephen
  Judd}, {and} \bibinfo{person}{Yevgeniy Vorobeychik}.}
  \bibinfo{year}{2012}\natexlab{}.
\newblock \showarticletitle{Behavioral experiments on a network formation
  game}. In \bibinfo{booktitle}{{\em Proceedings of the 13th ACM Conference on
  Electronic Commerce}}. ACM, \bibinfo{pages}{690--704}.
\newblock


\bibitem[\protect\citeauthoryear{LeBlanc and Koutsoukos}{LeBlanc and
  Koutsoukos}{2012}]%
        {Leblanc12}
\bibfield{author}{\bibinfo{person}{Heath~J. LeBlanc} {and}
  \bibinfo{person}{Xenofon~D. Koutsoukos}.} \bibinfo{year}{2012}\natexlab{}.
\newblock \showarticletitle{Low Complexity Resilient Consensus in Networked
  Multi-agent Systems with Adversaries}. In \bibinfo{booktitle}{{\em
  Proceedings of the 15th ACM International Conference on Hybrid Systems:
  Computation and Control}} {\em (\bibinfo{series}{HSCC '12})}.
  \bibinfo{publisher}{ACM}, \bibinfo{address}{New York, NY, USA},
  \bibinfo{pages}{5--14}.
\newblock
\showISBNx{978-1-4503-1220-2}
\showDOI{%
\url{http://dx.doi.org/10.1145/2185632.2185637}}


\bibitem[\protect\citeauthoryear{LeBlanc, Zhang, Koutsoukos, and
  Sundaram}{LeBlanc et~al\mbox{.}}{2013}]%
        {Leblanc13}
\bibfield{author}{\bibinfo{person}{H.~J. LeBlanc}, \bibinfo{person}{H. Zhang},
  \bibinfo{person}{X. Koutsoukos}, {and} \bibinfo{person}{S. Sundaram}.}
  \bibinfo{year}{2013}\natexlab{}.
\newblock \showarticletitle{Resilient Asymptotic Consensus in Robust Networks}.
\newblock \bibinfo{journal}{{\em IEEE Journal on Selected Areas in
  Communications\/}} \bibinfo{volume}{31}, \bibinfo{number}{4}
  (\bibinfo{date}{April} \bibinfo{year}{2013}), \bibinfo{pages}{766--781}.
\newblock
\showISSN{0733-8716}
\showDOI{%
\url{http://dx.doi.org/10.1109/JSAC.2013.130413}}


\bibitem[\protect\citeauthoryear{Leibbrandt, Ramalingam, Sääksvuori, and
  Walker}{Leibbrandt et~al\mbox{.}}{2015}]%
        {Leibbrandt15}
\bibfield{author}{\bibinfo{person}{Andreas Leibbrandt},
  \bibinfo{person}{Abhijit Ramalingam}, \bibinfo{person}{Lauri Sääksvuori},
  {and} \bibinfo{person}{James~M. Walker}.} \bibinfo{year}{2015}\natexlab{}.
\newblock \showarticletitle{Incomplete punishment networks in public goods
  games: experimental evidence}.
\newblock \bibinfo{journal}{{\em Experimental Economics\/}}
  \bibinfo{volume}{18}, \bibinfo{number}{1} (\bibinfo{year}{2015}),
  \bibinfo{pages}{15--37}.
\newblock


\bibitem[\protect\citeauthoryear{Mason and Suri}{Mason and Suri}{2012}]%
        {mason2012conducting}
\bibfield{author}{\bibinfo{person}{Winter Mason} {and}
  \bibinfo{person}{Siddharth Suri}.} \bibinfo{year}{2012}\natexlab{}.
\newblock \showarticletitle{Conducting behavioral research on Amazon’s
  Mechanical Turk}.
\newblock \bibinfo{journal}{{\em Behavior research methods\/}}
  \bibinfo{volume}{44}, \bibinfo{number}{1} (\bibinfo{year}{2012}),
  \bibinfo{pages}{1--23}.
\newblock


\bibitem[\protect\citeauthoryear{Matthew, Ramamohan, and Nicholas}{Matthew
  et~al\mbox{.}}{2009}]%
        {Mccubbins09}
\bibfield{author}{\bibinfo{person}{McCubbins Matthew}, \bibinfo{person}{Paturi
  Ramamohan}, {and} \bibinfo{person}{Weller Nicholas}.}
  \bibinfo{year}{2009}\natexlab{}.
\newblock \showarticletitle{Networked coordination: effect of network structure
  on human subjects' ability to solve coordination problem}.
\newblock \bibinfo{journal}{{\em Am Polit Res\/}}  \bibinfo{volume}{37}
  (\bibinfo{year}{2009}), \bibinfo{pages}{899--920}.
\newblock


\bibitem[\protect\citeauthoryear{Miller and Moser}{Miller and Moser}{2004}]%
        {Miller2004}
\bibfield{author}{\bibinfo{person}{John~H. Miller} {and} \bibinfo{person}{Scott
  Moser}.} \bibinfo{year}{2004}\natexlab{}.
\newblock \showarticletitle{{Communication and coordination}}.
\newblock \bibinfo{journal}{{\em Complexity\/}} \bibinfo{volume}{9},
  \bibinfo{number}{5} (\bibinfo{year}{2004}), \bibinfo{pages}{31--40}.
\newblock


\bibitem[\protect\citeauthoryear{Nguyen, Yang, Azaria, Kraus, and Tambe}{Nguyen
  et~al\mbox{.}}{2013}]%
        {Nguyen13}
\bibfield{author}{\bibinfo{person}{Thanh~Hong Nguyen}, \bibinfo{person}{Rong
  Yang}, \bibinfo{person}{Amos Azaria}, \bibinfo{person}{Sarit Kraus}, {and}
  \bibinfo{person}{Milind Tambe}.} \bibinfo{year}{2013}\natexlab{}.
\newblock \showarticletitle{Analyzing the Effectiveness of Adversary Modeling
  in Security Games}. In \bibinfo{booktitle}{{\em AAAI Conference on Artificial
  Intelligence}}. \bibinfo{pages}{718--724}.
\newblock


\bibitem[\protect\citeauthoryear{Olmstead, Viswanathan, Aicher, and
  Fowler}{Olmstead et~al\mbox{.}}{2009}]%
        {Olmstead2009}
\bibfield{author}{\bibinfo{person}{A.~J. Olmstead}, \bibinfo{person}{N.
  Viswanathan}, \bibinfo{person}{K.~A. Aicher}, {and} \bibinfo{person}{C.A.
  Fowler}.} \bibinfo{year}{2009}\natexlab{}.
\newblock \showarticletitle{{Sentence comprehension affects the dynamics of
  bimanual coordination: Implications for embodied cognition}}.
\newblock \bibinfo{journal}{{\em The Quarterly Journal of Experimental
  Psychology\/}}  \bibinfo{volume}{62} (\bibinfo{year}{2009}),
  \bibinfo{pages}{2409--2417}.
\newblock


\bibitem[\protect\citeauthoryear{Paolacci, Chandler, and Ipeirotis}{Paolacci
  et~al\mbox{.}}{2010}]%
        {paolacci2010running}
\bibfield{author}{\bibinfo{person}{Gabriele Paolacci}, \bibinfo{person}{Jesse
  Chandler}, {and} \bibinfo{person}{Panagiotis~G Ipeirotis}.}
  \bibinfo{year}{2010}\natexlab{}.
\newblock \showarticletitle{Running experiments on Amazon Mechanical Turk}.
\newblock \bibinfo{journal}{{\em Judgment and Decision Making\/}}
  \bibinfo{volume}{5}, \bibinfo{number}{5} (\bibinfo{year}{2010}).
\newblock


\bibitem[\protect\citeauthoryear{Peled, Kraus, et~al\mbox{.}}{Peled
  et~al\mbox{.}}{2015}]%
        {peled2015study}
\bibfield{author}{\bibinfo{person}{Noam Peled}, \bibinfo{person}{Sarit Kraus},
  {and} \bibinfo{person}{others}.} \bibinfo{year}{2015}\natexlab{}.
\newblock \showarticletitle{A study of computational and human strategies in
  revelation games}.
\newblock \bibinfo{journal}{{\em Autonomous Agents and Multi-Agent Systems\/}}
  \bibinfo{volume}{29}, \bibinfo{number}{1} (\bibinfo{year}{2015}),
  \bibinfo{pages}{73--97}.
\newblock


\bibitem[\protect\citeauthoryear{Pita, Jain, Ordonez, Tambe, and Kraus}{Pita
  et~al\mbox{.}}{2010}]%
        {Pita10}
\bibfield{author}{\bibinfo{person}{James Pita}, \bibinfo{person}{Manish Jain},
  \bibinfo{person}{Fernando Ordonez}, \bibinfo{person}{Milind Tambe}, {and}
  \bibinfo{person}{Sarit Kraus}.} \bibinfo{year}{2010}\natexlab{}.
\newblock \showarticletitle{Robust Solutions to Stackelberg Games: Addressing
  Bounded Rationality and Limited Observations in Human Cognition}.
\newblock \bibinfo{journal}{{\em Artificial Intelligence Journal\/}}
  \bibinfo{volume}{174}, \bibinfo{number}{15} (\bibinfo{year}{2010}),
  \bibinfo{pages}{1142--1171}.
\newblock


\bibitem[\protect\citeauthoryear{Rao and Monroe}{Rao and Monroe}{1989}]%
        {rao1989effect}
\bibfield{author}{\bibinfo{person}{Akshay~R Rao} {and} \bibinfo{person}{Kent~B
  Monroe}.} \bibinfo{year}{1989}\natexlab{}.
\newblock \showarticletitle{The effect of price, brand name, and store name on
  buyers' perceptions of product quality: An integrative review}.
\newblock \bibinfo{journal}{{\em Journal of marketing Research\/}}
  (\bibinfo{year}{1989}), \bibinfo{pages}{351--357}.
\newblock


\bibitem[\protect\citeauthoryear{Richerson and Boyd}{Richerson and
  Boyd}{2010}]%
        {Richerson2010}
\bibfield{author}{\bibinfo{person}{Peter~J. Richerson} {and}
  \bibinfo{person}{Robert Boyd}.} \bibinfo{year}{2010}\natexlab{}.
\newblock \showarticletitle{{Why possibly language evolved}}.
\newblock \bibinfo{journal}{{\em Biolinguistics\/}} \bibinfo{volume}{4},
  \bibinfo{number}{2-3} (\bibinfo{year}{2010}), \bibinfo{pages}{289--306}.
\newblock


\bibitem[\protect\citeauthoryear{Szamado}{Szamado}{2011}]%
        {Szamado2011}
\bibfield{author}{\bibinfo{person}{Szabolcs Szamado}.}
  \bibinfo{year}{2011}\natexlab{}.
\newblock \showarticletitle{{Pre-hunt communication provides context for the
  evolution of early human languge}}.
\newblock \bibinfo{journal}{{\em Biological Theory\/}} \bibinfo{volume}{5},
  \bibinfo{number}{4} (\bibinfo{year}{2011}), \bibinfo{pages}{366--382}.
\newblock


\bibitem[\protect\citeauthoryear{Thaler and Sunstein}{Thaler and
  Sunstein}{2009}]%
        {Thaler09}
\bibfield{author}{\bibinfo{person}{Richard~H. Thaler} {and}
  \bibinfo{person}{Cass~R. Sunstein}.} \bibinfo{year}{2009}\natexlab{}.
\newblock \bibinfo{booktitle}{{\em Nudge: Improving Decisions About Health,
  Wealth, and Happiness}}.
\newblock \bibinfo{publisher}{Penguin Books}.
\newblock


\bibitem[\protect\citeauthoryear{Vorobeychik, Joveski, and Yu}{Vorobeychik
  et~al\mbox{.}}{2017}]%
        {vorobeychik2017does}
\bibfield{author}{\bibinfo{person}{Yevgeniy Vorobeychik},
  \bibinfo{person}{Zlatko Joveski}, {and} \bibinfo{person}{Sixie Yu}.}
  \bibinfo{year}{2017}\natexlab{}.
\newblock \showarticletitle{Does communication help people coordinate?}
\newblock \bibinfo{journal}{{\em PloS One\/}} \bibinfo{volume}{12},
  \bibinfo{number}{2} (\bibinfo{year}{2017}), \bibinfo{pages}{1--19}.
\newblock


\bibitem[\protect\citeauthoryear{Watts and Strogatz}{Watts and
  Strogatz}{1998}]%
        {watts1998collective}
\bibfield{author}{\bibinfo{person}{Duncan~J Watts} {and}
  \bibinfo{person}{Steven~H Strogatz}.} \bibinfo{year}{1998}\natexlab{}.
\newblock \showarticletitle{Collective dynamics of ‘small-world’networks}.
\newblock \bibinfo{journal}{{\em nature\/}} \bibinfo{volume}{393},
  \bibinfo{number}{6684} (\bibinfo{year}{1998}), \bibinfo{pages}{440}.
\newblock


\bibitem[\protect\citeauthoryear{Wunder, Suri, and Watts}{Wunder
  et~al\mbox{.}}{2013}]%
        {Wunder13}
\bibfield{author}{\bibinfo{person}{Michael Wunder}, \bibinfo{person}{Siddharth
  Suri}, {and} \bibinfo{person}{Duncan~J Watts}.}
  \bibinfo{year}{2013}\natexlab{}.
\newblock \showarticletitle{Empirical agent based models of cooperation in
  public goods games}. In \bibinfo{booktitle}{{\em Proceedings of the
  fourteenth ACM conference on Electronic commerce}}. ACM,
  \bibinfo{pages}{891--908}.
\newblock


\bibitem[\protect\citeauthoryear{Zeng and Chow}{Zeng and Chow}{2014}]%
        {zeng2014resilient}
\bibfield{author}{\bibinfo{person}{Wente Zeng} {and} \bibinfo{person}{Mo-Yuen
  Chow}.} \bibinfo{year}{2014}\natexlab{}.
\newblock \showarticletitle{Resilient distributed control in the presence of
  misbehaving agents in networked control systems}.
\newblock \bibinfo{journal}{{\em IEEE transactions on cybernetics\/}}
  \bibinfo{volume}{44}, \bibinfo{number}{11} (\bibinfo{year}{2014}),
  \bibinfo{pages}{2038--2049}.
\newblock


\bibitem[\protect\citeauthoryear{Zhang, Vorobeychik, Letchford, and
  Lakkaraju}{Zhang et~al\mbox{.}}{2016}]%
        {Zhang16}
\bibfield{author}{\bibinfo{person}{Haifeng Zhang}, \bibinfo{person}{Yevgeniy
  Vorobeychik}, \bibinfo{person}{Joshua Letchford}, {and}
  \bibinfo{person}{Kiran Lakkaraju}.} \bibinfo{year}{2016}\natexlab{}.
\newblock \showarticletitle{Data-driven agent-based modeling, with application
  to rooftop solar adoption}.
\newblock \bibinfo{journal}{{\em Journal of Autonomous Agents and Multiagent
  Systems\/}} \bibinfo{volume}{30}, \bibinfo{number}{6} (\bibinfo{year}{2016}),
  \bibinfo{pages}{1023--1049}.
\newblock


\end{thebibliography}

\end{document}